\documentclass[12pt]{article}

\usepackage{epsfig}

\usepackage{graphicx}

\def\square{\kern1pt\vbox{\hrule height 1.2pt
\hbox{\vrule width 1.2pt\hskip 3pt
\vbox{\vskip 6pt}\hskip 3pt\vrule width 0.6pt}
\hrule height 0.6pt}\kern1pt}
\def\ltwid{\mathrel{\raise.3ex\hbox{$<$\kern-.75em\lower1ex\hbox{$\sim$}}}}

\def\ltwid{\mathrel{\raise.3ex\hbox{$<$\kern-.75em\lower1ex\hbox{$\sim$}}}}
\def\gtwid{\mathrel{\raise.3ex\hbox{$>$\kern-.75em\lower1ex\hbox{$\sim$}}}}

\begin{document}

\begin{titlepage}
\begin{flushright}
CCTP-2010-8 \\ UFIFT-QG-10-04
\end{flushright}

\vspace{0.5cm}

\begin{center}
\bf{Primordial Density Perturbations and Reheating from Gravity}
\end{center}

\vspace{0.3cm}

\begin{center}
N. C. Tsamis$^{\dagger}$
\end{center}
\begin{center}
\it{Department of Physics, University of Crete \\
GR-710 03 Heraklion, HELLAS.}
\end{center}

\vspace{0.2cm}

\begin{center}
R. P. Woodard$^{\ast}$
\end{center}
\begin{center}
\it{Department of Physics, University of Florida \\
Gainesville, FL 32611, UNITED STATES.}
\end{center}

\vspace{0.3cm}

\begin{center}
ABSTRACT
\end{center}
\hspace{0.3cm} We consider the presence and evolution
of primordial density perturbations in a cosmological
model based on a simple {\it ansatz} which captures --
by providing a set of effective gravitational field 
equations -- the strength of the enhanced quantum loop 
effects that can arise during inflation. After deriving
the general equations that perturbations obey, we 
concentrate on scalar perturbations and show that their
evolution is quite different than that of conventional
inflationary models but still phenomenologically
acceptable. The main reason for this novel evolution
is the presence of an oscillating regime after the end
of inflation which makes {\it all} super-horizon scalar
modes oscillate. The same reason allows for a natural
and very fast reheating mechanism for the universe.

\vspace{0.3cm}

\begin{flushleft}
\end{flushleft}

\vspace{0.1cm}

\begin{flushleft}
$^{\dagger}$ e-mail: tsamis@physics.uoc.gr \\
$^{\ast}$ e-mail: woodard@phys.ufl.edu
\end{flushleft}

\end{titlepage}

\section{Introduction}

During the inflationary era infrared gravitons are 
produced out of the vacuum because of the accelerated
expansion of spacetime. Such a production can only occur
for particles that are light compared to the Hubble scale
without classical conformal invariance; gravitons and
massless minimally coupled scalars are unique in that
respect.

The self-gravitation of the vast ensemble of inflationary 
gravitons must act to slow the expansion rate \cite{NctRpw1}. 
This effect is inherently non-local because it couples the 
local graviton energy density with the potential induced 
by the interaction stress of the gravitons throughout the 
past light-cone. This suggests that the relevant effective 
field equations should be non-local. Non-local models of 
cosmology have been much studied because they can avoid 
the problem that de Sitter must be a solution for any 
local, stable theory, and because non-local couplings 
between different times can ease fine tuning problems 
\cite{ours,theirs}.

Quantum gravitational loop corrections which can be 
computed during inflation grow like the logarithm of 
the inflationary scale factor \cite{TW1,SW,MW1,many}. 
It should ultimately be possible to derive the most 
cosmologically significant part of the effective field 
equations by summing the series of leading infrared 
logarithms. Starobinsky has proposed a technique for 
accomplishing this \cite{AAS}, and Starobinsky and 
Yokoyama have applied it to scalar potential models 
\cite{SY}. Starobinsky's method has recently been 
extended to Yukawa-coupled fermions \cite{MW2} and 
to scalar quantum electrodynamics \cite{PTsW}. It 
has not yet been extended to quantum gravity but 
there are reasons for believing that some version 
of it can be \cite{MW3}.

Although full control of the effect requires a 
non-perturbative resummation technique, one can 
attempt to anticipate the results of such a formalism
in a variety of ways. One approach is to simplify the 
full quantum gravitational dynamics by assuming that 
the exact graviton remains transverse-traceless and 
free, but propagates in the background geometry of 
an effective scale factor which is determined from 
the expectation value of the $g_{00}$ gravitational 
constraint equation of motion \cite{NctRpw2}. The 
simplified theory retains the proper perturbative limit
for de Sitter spacetime and may provide the basis for a 
tractable non-perturbative formulation.

Another approach is to use the physical principles responsible 
for the non-trivial quantum gravitational back-reaction on
inflation and construct a phenomenological model which we
can then directly evolve. In a previous paper \cite{NctRpw3} 
we proposed a phenomenological model which can provide 
evolution beyond perturbation theory. In one sentence, we 
constructed an {\it effective} conserved stress-energy 
tensor $T_{\mu\nu} [g]$ which modifies the gravitational 
equations of motion:
\footnote{Hellenic indices take on spacetime values while 
Latin indices take on space values. Our metric tensor 
$g_{\mu\nu}$ has spacelike signature and our curvature 
tensor equals:
$R^{\alpha}_{~\beta\mu\nu} \equiv 
\Gamma^{\alpha}_{~\nu\beta, \mu} +
\Gamma^{\alpha}_{~\mu\rho} \;
\Gamma^{\rho}_{~\nu\beta} -
(\mu \leftrightarrow \nu)$. 
The initial value of the Hubble parameter is 
$3H^2_0 \equiv \Lambda$. 
We restrict our analysis to scales 
$M \equiv (\, \Lambda / 8 \pi G \,)^{\frac14}$ 
below the Planck mass $M_{\rm Pl} \equiv G^{-1}$ 
so that the dimensionless coupling constant 
$\epsilon \equiv G \Lambda$ of the theory is 
small.}
\begin{equation}
G_{\mu\nu} \; \equiv \;
R_{\mu\nu} \, - \, \frac12 g_{\mu\nu} \, R \; = \;
- \Lambda \, g_{\mu\nu} \, + \,
8 \pi G \, T_{\mu\nu}[g] 
\;\; . \label{eom1}
\end{equation}
and which, we hope, contains the most cosmologically
significant part of the full effective quantum
gravitational equations.

Our physical {\it ansatz} consists of parametrizing
$T_{\mu\nu}[g]$ as a ``perfect fluid'':
\begin{equation}
T_{\mu\nu}[g] \; = \;
(\rho + p) \, u_{\mu} \, u_{\nu} \, + \,
p \, g_{\mu\nu}
\;\; , \label{Tmn}
\end{equation}
with the gravitationally induced pressure given
as the following functional of the metric tensor:
\begin{equation}
p[g](x) \; = \;
\Lambda^2 \, f[- \epsilon \, X](x) 
\qquad , \qquad
X \, \equiv \, \frac{1}{\square} \, R
\;\; , \label{p}
\end{equation}
where the function $f$ satisfies:
\begin{equation}
f[- \epsilon \, X] \; = \;
- \epsilon \, X \, + \, O(\epsilon^2)
\;\; , \label{fincr}
\end{equation}
and where the scalar d'Alembertian:
\begin{equation}
\square \, \equiv \;
\frac{1}{\sqrt{-g}} \;
\partial_{\mu} \Big( \,
g^{\mu\nu} \sqrt{-g} \; \partial_{\nu} \, \Big)
\;\; , \label{box}
\end{equation}
is defined with retarded boundary conditions.
The induced energy density $\rho[g]$ and 4-velocity
$u_{\mu}[g]$ are determined, up to their initial
value data, from stress-energy conservation:
\begin{equation}
D^{\mu} \, T_{\mu\nu} \; = \; 0
\;\; , \label{cons1}
\end{equation}
which implies:
\begin{eqnarray}
& \mbox{} &
u \cdot \partial \, p \; = \;
D_{\mu} \Big[ \, (\rho + p) u^{\mu} \, \Big]
\;\; , \label{cons2a} \\
& \mbox{} &
u \cdot \partial \, \rho \; = \; 
- (D \cdot u) (\rho + p)
\;\; , \label{cons2b} \\
& \mbox{} &
(\rho + p) \; u \cdot D u_{\nu} \; = \;
- ( \, \partial_{\nu} \, + \, 
u_{\nu} \, u \cdot \partial \, ) \, p
\;\; . \label{cons2c}
\end{eqnarray}
The 4-velocity is chosen to be timelike and
normalized:
\begin{equation}
g^{\mu\nu} \, u_{\mu} u_{\nu} = -1
\qquad \Longrightarrow \qquad
u^{\mu} \, u_{\mu ; \nu} = 0
\;\; . \label{u}
\end{equation}

The purpose of this paper is to study the behaviour
of the primordial scalar perturbations in the
phenomenological model summarized above. Due to
the non-local structure of the model the standard
perturbation analysis must be extended and this 
is done in Section 3 after the relevant cosmological
background is presented in Section 2. The evolution
and novel features of the scalar perturbations are
addressed in Section 4 and their normalization in
Section 5. A result of the novel behaviour of our
scalar mode functions after the end of the inflationary
era allows a natural and very fast mechanism for
reheating the universe which is described in Section 
6. Our conclusions comprise Section 7. \\

\section{The Cosmological Background}

The large-scale homogeneity and isotropy of the universe
selects Friedman-Robertson-Walker ($FRW$) spacetimes as 
those of primary cosmological interest; their line element
for zero spatial curvature equals in co-moving coordinates:
\begin{equation}
d{\bar s}^2 \; = \; 
{\bar g}_{\mu\nu}(t) \; dx^{\mu} dx^{\nu} \; = \; 
- dt^2 \, + \, a^2(t) \; d{\bf x} \cdot d{\bf x} 
\;\; . \label{ds^2frw}
\end{equation}   
Derivatives of the scale factor $a(t)$ give the Hubble 
parameter $H(t)$ -- a measure of the cosmic expansion rate 
-- and the deceleration parameter $q(t)$ -- a measure of 
the cosmic acceleration:
\begin{eqnarray}
H(t) & \equiv &
\frac{{\dot a}(t)}{a(t)} \, = \,
\frac{d}{dt} \ln a(t)
\;\; , \label{H} \\
q(t) & \equiv &
- \frac{a(t) \, {\ddot a}(t)}{{\dot a}^2(t)} \, = \,
-1 \, - \, \frac{\dot H(t)}{H^2(t)}
\;\; . \label{q}
\end{eqnarray} 

For these spacetimes the stress-energy tensor (\ref{Tmn}) 
equals: 
\begin{eqnarray}
{\bar T}_{00} & = & 
{\bar u}_0 \, {\bar u}_0 \,
( {\bar \rho} + {\bar p} ) \, - \, 
{\bar p} \; = \;
{\bar \rho}
\;\; , \label{T00frw} \\
{\bar T}_{0i} & = & 0
\;\; , \label{T0ifrw} \\
{\bar T}_{ij} & = & 
{\bar u}_i \, {\bar u}_j \,
( {\bar \rho} + {\bar p} ) \, + \, 
{\bar g}_{ij} \, {\bar p} \; = \;
{\bar g}_{ij} \, {\bar p}
\;\; . \label{Tijfrw}
\end{eqnarray} 
An immediate consequence of isotropy and the normalization
condition (\ref{u}) is: 
\begin{equation}
{\bar u}_{\mu} \; = \;
- \, \delta_{\mu}^{~0}
\qquad \Longleftrightarrow \qquad
{\bar u}^{\mu} \; = \;
\delta^{\mu}_{~0}
\;\; . \label{umfrw}
\end{equation}
The Ricci tensor and Ricci scalar become, respectively:
\begin{eqnarray}
{\bar R}_{00} & = & 
- \left[ \, \frac{3{\ddot a}}{a} \, \right] \; = \; 
- \left( \, 3H^2 \, + \, 3 {\dot H} \, \right)
\;\; , \label{R00frw} \\
{\bar R}_{0i} & = & 
0
\;\; , \label{R0ifrw} \\
{\bar R}_{ij} & = & 
\left[ \, \frac{{\ddot a}}{a} \, + \,
\frac{2{\dot a}^2}{a^2} \, \right]  
{\bar g}_{ij} \; = \; 
\left( \, 3H^2 \, + \, {\dot H} \, \right)
{\bar g}_{ij}
\;\; , \label{Rijfrw} 
\end{eqnarray}
and:
\begin{equation}
{\bar R} \; = \;
\left[ \, \frac{6{\ddot a}}{a} \, + \,
\frac{6{\dot a}^2}{a^2} \, \right] \; = \; 
\left( \, 12H^2 \, + \, 6 {\dot H} \, \right)
\;\; . \label{Rfrw}
\end{equation}

In view of (\ref{T00frw}-\ref{Tijfrw},
\ref{R00frw}-\ref{Rfrw}), the gravitational equations 
of motion (\ref{eom1}) take the form:
\begin{eqnarray}
3H^2 & = &
\Lambda \, + \, 8 \pi G \, {\bar \rho}
\;\; , \label{g00frw} \\
-2 {\dot H} - 3 H^2 & = & \!
- \Lambda \, + \, 8 \pi G \, {\bar p}
\;\; ,\label{gijfrw}
\end{eqnarray}
while the conservation equation (\ref{cons1}) becomes:
\begin{equation}
{\dot{\bar \rho}} \; = \;
- 3H \, ( {\bar \rho} + {\bar p} ) 
\;\; . \label{consfrw}
\end{equation}
The latter implies that:
\begin{equation}
{\bar \rho}(t) \; = \; 
- {\bar p}(t) \, + \,
\frac{1}{a^3(t)} 
\int_0^t dt' \; a^3(t') \; {\dot {\bar p}}(t') 
\;\; . \label{rhofrw-pfrw} \\
\end{equation}

When acting on functions which only depend on co-moving
time, the scalar d'Alembertian (\ref{box}) for $FRW$ 
geometries equals:
\begin{equation}
\bar{\square} \; = \;
- \left( \, \partial_t^2 \, + \, 3H \partial_t \, \right)
\;\; , \label{boxfrw}
\end{equation}
so that its inverse is:
\begin{equation}
\frac{1}{\bar{\square}} \; = \;
- \int_0^t dt' \; \frac{1}{a^3(t')} \, 
\int_0^{t'} dt'' \; a^3(t'')
\;\; . \label{invboxfrw} 
\end{equation}
Consequently, the source ${\bar X}$ can be written as
follows:
\begin{equation}
{\bar X} \; = \;
\frac{1}{\bar{\square}} \, {\bar R} \; = \;
- \int_0^t dt' \; \frac{1}{a^3(t')} \,
\int_0^{t'} dt'' \; a^3(t'') \,
\left[ \, 12 H^2(t'') \, + \, 6 {\dot H}^2(t'') \, \right]
\;\; . \label{Xfrw} 
\end{equation}
Note that we have taken the initial time to be at 
$t=0$.

\section{Perturbations}

Small deviations from the homogeneity and isotropy of the 
$FRW$ geometries are necessary to address, among other 
issues, the existence of primordial density perturbations. 
To account for these deviations, we first define small but 
general perturbations of all relevant variables about 
their $FRW$ values. In co-moving coordinates we have:
\begin{eqnarray}
p(t, {\bf x}) & \equiv & 
{\bar p}(t) + \Delta p(t, {\bf x})
\;\; , \label{pertp} \\
\rho(t, {\bf x}) & \equiv & 
{\bar \rho}(t) \, + \, \Delta \rho(t, {\bf x})
\;\; , \label{pertrho} \\
u_{0}(t, {\bf x}) & \equiv & 
{\bar u}_{0} \, + \, 
\Delta u_{0}(t, {\bf x})
\;\; , \label{pertu0} \\
u_{i}(t, {\bf x}) & \equiv &
{\bar u}_{i} \, + \, 
a(t) \, \Delta u_{i}(t, {\bf x})
\;\; , \label{pertui} \\
g_{00}(t, {\bf x}) & \equiv & 
{\bar g}_{00} \, + \, h_{00}(t, {\bf x})
\; = \;
-1 \, + \, h_{00}(t, {\bf x})
\;\; , \label{pertg00} \\
g_{0i}(t, {\bf x}) & \equiv &  
{\bar g}_{0i} \, + \,a(t) \, h_{0i}(t, {\bf x})
\; = \;
a(t) \, h_{0i}(t, {\bf x})
\;\; , \label{pertg0i} \\
g_{ij}(t, {\bf x}) & \equiv &
{\bar g}_{ij}(t) \, + \, 
a^2(t) \, h_{ij}(t, {\bf x})
\; = \;
a^2(t) \, [ \, \delta_{ij} +  h_{ij}(t, {\bf x}) \, ]
\;\; . \label{pertgij}
\end{eqnarray}
We then substitute (\ref{pertp}-\ref{pertgij}) in the 
gravitational equations of motion (\ref{eom1}):
\begin{eqnarray}
G^{\mu}_{~\nu} \!\!& \equiv &\!\!
R^{\mu}_{~\nu} \, - \, \frac12 \, \delta^{\mu}_{~\nu} \, R 
\nonumber \\
\!\!& = &\!\! 
- \Lambda \, \delta^{\mu}_{~\nu} \, + \,
8\pi G \, \Big[ \, (\rho + p) \, u^{\mu} \, u_{\nu} \, + \,
p \, \delta^{\mu}_{~\nu} \, \Big]
\; \equiv \;
- \Lambda \, \delta^{\mu}_{~\nu} \, + \,
T^{\mu}_{~\nu}
\;\; , \qquad \label{eom3}
\end{eqnarray}
and obtain equations to the desired order in the
perturbation; for our purposes, the first order
equations will suffice. This is a tedious process 
which to some degree can be simplified by a proper 
choice of coordinate system and field variables. 
It turns out that conformal coordinates 
$(\eta, {\bf x})$ and a set of gauge invariant 
variables is the optimal choice \cite{mukhanov1}. 

In conformal coordinates -- which we shall use 
thereafter in this Section -- the background 
invariant element is proportional to that of 
flat spacetime:
\begin{equation}
d{\bar s}^2 \; = \; 
{\bar g}_{\mu\nu}(\eta) \; dx^{\mu} dx^{\nu} \; = \; 
a^2(\eta) \, \Big( \! -d\eta^2 \, + \, 
d{\bf x} \cdot d{\bf x} \Big)
\;\; . \label{ds^2frwconf} \\
\end{equation}   
The relation between co-moving and conformal times is:
\begin{equation}
d\eta \, = \, \frac{dt}{a(t)} 
\qquad \Longrightarrow \qquad
\frac{d}{d\eta} \, = \, a \, \frac{d}{dt}
\;\; . \label{conftime} \\
\end{equation}   
The corresponding perturbations take the form:
\begin{eqnarray}
p(\eta, {\bf x}) & \equiv & 
{\bar p}(\eta) + \Delta p(\eta, {\bf x})
\;\; , \label{pertp2} \\
\rho(\eta, {\bf x}) & \equiv & 
{\bar \rho}(\eta) \, + \, \Delta \rho(\eta, {\bf x})
\;\; , \label{pertrho2} \\
u_{\mu}(\eta, {\bf x}) & \equiv &
{\bar u}_{\mu} (\eta) \, + \, 
a(\eta) \, \Delta u_{\mu}(\eta, {\bf x})
\; = \;
a(\eta) \, [ \, - \delta_{\mu}^{~0} \, + \, 
\Delta u_{\mu}(\eta, {\bf x}) \, ]
\;\; , \qquad \label{pertu} \\
g_{\mu\nu}(\eta, {\bf x}) & \equiv &
{\bar g}_{\mu\nu}(\eta) \, + \,
a^2(\eta) \, h_{\mu\nu}(t, {\bf x})
\; = \;
a^2(\eta) \, [ \, 
\eta_{\mu\nu} +  h_{\mu\nu}(t, {\bf x}) \, ]
\;\; , \label{pertgmn}
\end{eqnarray}
where we have used the conformal analogue of
(\ref{umfrw}):
\begin{equation}
{\bar u}_{\mu} \; = \;
- a \, \delta_{\mu}^{~0}
\qquad \Longleftrightarrow \qquad
{\bar u}^{\mu} \; = \;
a^{-1} \, \delta^{\mu}_{~0}
\;\; . \label{baruconf}
\end{equation}

\vspace{0.3cm}

{$\bullet \;\; $} {\bf The Left Hand Side} \\
As a result of straightforward manipulations, 
the left hand side of the field equations 
(\ref{eom3}) can be written as:
\begin{equation}
G^{\mu}_{~\nu} \; \equiv \;
{\bar G}^{\mu}_{~\nu} +
\Delta G^{\mu}_{~\nu}
\;\; , \label{Gmnconf1}
\end{equation}
with:
\footnote{In conformal coordinates, the prime superscript 
denotes differentiation with respect to conformal time.}
\begin{eqnarray}
{\bar G}^{\mu}_{~\nu} 
\!\!& = &\!\!
\delta^{\mu}_{~0} \, 
\delta^{\nu}_{~0} 
\left[ \, 2 \, \frac{a''}{a^3} - 
4 \, \frac{a'^{\, 2}}{a^4} \, \right] 
\, + \,
\delta^{\mu}_{~\nu}  
\left[ \, -2 \, \frac{a''}{a^3} + 
\frac{a'^{\, 2}}{a^4} \, \right] 
\;\; , \label{Gmnbarconf1} \\
{\Delta G}^{\mu}_{~\nu} 
\!\!& = &\!\!
- \, \delta^0_{~\nu} \, h^{\mu\nu} 
\left[ \, -2 \, \frac{a''}{a^3} + 
\frac{a'^{\, 2}}{a^4} \, \right]
\, + \,
\delta^{\mu}_{~\nu} \, h^{00}  
\left[ \, -2 \, \frac{a''}{a^3} + 
\frac{a'^{\, 2}}{a^4} \, \right] 
\nonumber \\
& \mbox{} &
- \left( \, \eta^{\mu\rho} \, \delta^{\sigma}_{~\nu} -
\delta^{\mu}_{~\nu} \, \eta^{\rho\sigma} \, \right)
\left[ \, h_{0\rho , \sigma} + h_{0\sigma , \rho} -
h'_{\rho\sigma} \, \right] 
\frac{a'}{a^3}
\nonumber \\
& \mbox{} &
+ \left[ \, 
h^{\mu\rho}_{~~, \rho\nu} +
h_{\nu\rho}^{~~, \rho\mu} -
h^{\mu ~, \rho}_{~\nu ~~ \rho} -
h^{\rho ~, \mu}_{~\rho ~~ \nu} 
\, \right] \frac{1}{2a^2}
\nonumber \\
& \mbox{} &
- \, \delta^{\mu}_{~\nu} 
\left[ \, 
h^{\rho\sigma}_{~~, \rho\sigma} -
h^{\rho ~, \sigma}_{~\rho ~~ \sigma}
\, \right] \frac{1}{2a^2}
\;\; . \label{deltaGmnconf1}
\end{eqnarray}
It is convenient to 3+1 decompose the background value 
${\bar G}^{\mu}_{~\nu}$ in (\ref{Gmnbarconf1}): 
\begin{eqnarray}
{\bar G}^0_{~0} \!\!& = &\!\!
-3 \, \frac{a'^{\, 2}}{a^4}
\;\; , \label{barG00conf1} \\
{\bar G}^0_{~i} \!\!& = &\!\!
0
\;\; , \label{barG0iconf1} \\
{\bar G}^i_{~j} \!\!& = &\!\!
\left[ \, -2 \, \frac{a''}{a^3} + 
\frac{a'^{\, 2}}{a^4} \, \right] 
\delta_{ij}
\;\; , \label{barGijconf1}
\end{eqnarray}
as well as the first order perturbation 
${\Delta G}^{\mu}_{~\nu}$ in (\ref{deltaGmnconf1}):
\begin{eqnarray}
{\Delta G}^0_{~0} 
\!\!& = &\!\!
-3 \, \frac{a'^{\, 2}}{a^4} \, h_{00} 
\, + \,
\frac{a'}{a^3} \, \Big[ \,
2 h_{0i,i} - h'_{ii} \, \Big]
\, + \, 
\frac{1}{2a^2} \left[ \,
\nabla^2 h_{ii} - h_{ij,ij} \, \right]
\;\; , \label{deltaG00conf1} \\
{\Delta G}^0_{~i} 
\!\!& = &\!\!
\frac{a'}{a^3} \, h_{00,i} 
\, + \,
\frac{1}{2a^2} \left[ \,
-h_{0j,ij} + \nabla^2 h_{0i} 
- h'_{ij,ij} + h'_{jj,i} \, \right]
\;\; , \label{deltaG0iconf1} \\
{\Delta G}^i_{~j} 
\!\!& = &\!\!
\delta_{ij} \, \Bigg\{ 
\left[ \, -2 \, \frac{a''}{a^3} + 
\frac{a'^{\, 2}}{a^4} \, \right] h_{00}
\, + \,
\frac{a'}{a^3} \, \Big[
-h'_{00} + 2 h_{0k,k} - h'_{kk} \, \Big]
\nonumber \\
& \mbox{} &
+ \, \frac{1}{2a^2} \, \Big[
-h''_{kk} + \nabla^2 h_{kk} + 2 h'_{0k,k} - 
h_{k\ell , k\ell} - \nabla^2 h_{00} \, \Big]
\Bigg\}
\nonumber \\
& \mbox{} &
+ \, \frac{a'}{a^3} \, \Big[
-2h_{0(i,j)} + h'_{ij} \, \Big]
\nonumber \\
& \mbox{} &
+ \, \frac{1}{a^2} \, \Big[ \,
h''_{ij} - \nabla^2 h_{ij} -
2h'_{0(i,j)} + 2h_{k(i,j)k} +
h_{00,ij} - h_{kk,ij} \, \Big]
\;\; . \qquad \label{deltaGijconf1}
\end{eqnarray}

\vspace{0.3cm}

{$\star \; $} {\it Identities} \\
In addition to the equations presented throughout the 
main text, various of the following expressions have 
been used to obtain the results of this subsection: 
\begin{eqnarray}
a_{, \rho} \!\!& = &\!\!
a' \, \delta^{0}_{~\rho}
\qquad , \qquad 
a_{, \rho\sigma} \; = \;
a'' \, \delta^{0}_{~\rho} \, \delta^{0}_{~\sigma}
\;\; , \label{IDa1}
\end{eqnarray}

\vspace{-0.9cm}

\begin{eqnarray}
\Gamma^{\rho}_{~\mu\nu} \!\!& \equiv &\!\!
\bar{\Gamma}^{\rho}_{~\mu\nu} \, + \,
\Delta \Gamma^{\rho}_{~\mu\nu} 
\;\; , \label{IDGamma1} \\
\bar{\Gamma}^{\rho}_{~\mu\nu} \!\!& = &\!\!
\frac{a'}{a} \, \Big[ \,
\delta^{\rho}_{~\mu} \, \delta^{0}_{~\nu} +
\delta^{\rho}_{~\nu} \, \delta^{0}_{~\mu} -
\eta^{\rho 0} \, \eta_{\mu\nu} \, \Big]
\;\; , \label{IDGamma2} \\ 
\Delta \Gamma^{\rho}_{~\mu\nu} \!\!& = &\!\!
\frac{a'}{a} \, \Big[ \,
\eta_{\mu\nu} \, h^{\rho 0} -
\eta^{\rho 0} \, h_{\mu\nu} \, \Big] \, + \,
\frac12 \, \Big[ \,
h^{\rho}_{~\mu , \nu} +
h^{\rho}_{~\nu , \mu} -
h_{\mu\nu}^{~~, \rho} \, \Big]
\;\; , \label{IDGamma3} 
\end{eqnarray}

\vspace{0.3cm}

{$\bullet \;\; $} {\bf The Right Hand Side}  \\
We write the gravitationally induced stress-energy 
tensor $T^{\mu}_{~\nu}$ as:
\begin{equation}
T^{\mu}_{~\nu} \; \equiv \;
{\bar T}^{\mu}_{~\nu} \, + \,
{\Delta T}^{\mu}_{~\nu} 
\;\; . \label{Tmnconf1}
\end{equation}
Its background value ${\bar T}^{\mu}_{~\nu}$ is:
\begin{equation}
{\bar T}^{\mu}_{~\nu} \; = \;
( {\bar \rho} + {\bar p} ) \, 
{\bar u}^{\mu} \, {\bar u}_{\nu} \, + \,
{\bar p} \, \delta^{\mu}_{~\nu}
\;\; , \label{barTmnconf1}
\end{equation}
and, using (\ref{baruconf}), its 3+1 decomposition 
takes the form:
\begin{equation}
 {\bar T}^0_{~0} \; = \; - {\bar \rho}
\qquad , \qquad 
{\bar T}^0_{~i} \; = \; 0
\qquad , \qquad 
{\bar T}^i_{~j} \; = \; {\bar p} \, \delta_{ij}
\;\; . \label{barTmnconf2}
\end{equation}

The perturbation ${\Delta T}^{\mu}_{~\nu}$ defined
in (\ref{Tmnconf1}) equals: 
\begin{equation}
{\Delta T}^{\mu}_{~\nu} \; = \;
{\bar u}^{\mu} \, {\bar u}_{\nu} \; {\Delta (\rho + p)} 
\, + \,
({\bar \rho} + {\bar p}) \; {\Delta (u^{\mu} \, u_{\nu})}
\, + \,
{\Delta p} \; \delta^{\mu}_{~\nu}
\;\; , \label{deltaTmnconf1} 
\end{equation}
and, in view of (\ref{baruconf}), the 3+1
decomposition is given by:
\begin{eqnarray}
{\Delta T}^{0}_{~0} \!\!& = &\!\!
- {\Delta \rho}  \, + \,
({\bar \rho} + {\bar p}) 
\Big[ -a \, \Delta u^0 + 
a^{-1} \, \Delta u_0 \, \Big]  
\;\; , \label{deltaT00conf1} \\
{\Delta T}^{0}_{~i} \!\!& = &\!\!
({\bar \rho} + {\bar p}) \, 
a^{-1} \, \Delta u_i 
\;\; , \label{deltaT0iconf1} \\
{\Delta T}^{i}_{~j} \!\!& = &\!\!
{\Delta p} \; \delta^{i}_{~j} 
\;\; . \label{deltaTijconf1} 
\end{eqnarray}

* {\it The induced pressure deviation $\Delta p$} \\
The starting point for the explicit computations 
is the induced pressure {\it ansatz} (\ref{p}) 
which we shall expand to first order about the 
background geometry and determine $\Delta p$. 
Knowledge of $\Delta p$ and use of the conservation
equations will allow us to obtain the remaining 
deviations $\Delta \rho$ and $\Delta u_{\mu}$,
up to initial value data. 
\footnote{As we shall see later in this subsection,
the perturbation $\Delta u_0$ can be computed 
independent of the conservation equations.}
The aforementioned expansion is:
\begin{equation}
p \; = \;
\Lambda^2 \, f[- \epsilon \, X] \; = \;
\Lambda^2 \, \Big\{ \, 
f[- \epsilon \, {\bar X}] \, - \,
f'[- \epsilon \, {\bar X}] \, (\epsilon \, \Delta X)
\, + \, O(\epsilon^2) \, \Big\}
\;\; , \label{pexp}
\end{equation}
where:
\begin{equation}
f'[ -\epsilon \, {\bar X} ] \; \equiv \;
- \frac{1}{\epsilon} \,
\frac{d}{d{\bar X}} f[ -\epsilon \, {\bar X} ] 
\;\; . \label{f'}
\end{equation}
Therefore:
\begin{eqnarray}
{\bar p} \!\!& = &\!\! 
\Lambda^2 \, f[- \epsilon \, {\bar X}] 
\qquad , \qquad 
{\bar X} \; = \; {\bar{\square}}^{-1} {\bar R}
\;\; , \label{barpconf1} \\
{\Delta p} \!\!& = &\!\! 
- \epsilon \Lambda^2 \, 
f'[- \epsilon \, {\bar X}] \; \Delta X
\;\; . \label{deltapconf1}
\end{eqnarray}
Now the first order perturbation $\Delta X$ equals:
\begin{equation}
\Delta X \; \equiv \;
\Delta (\square^{-1} R) \; = \;
\frac{1}{\bar{\square}} \left[ \,
\Delta R -  (\Delta \square) \, {\bar X} \, \right] 
\;\; , \label{deltaX}
\end{equation}
and we must evaluate: \\
{\it (i) the Ricci scalar perturbation} $\Delta R \! :$ 
\begin{equation}
R \; \equiv \;
{\bar R} \, + \, \Delta R
\;\; , \label{deltaR}
\end{equation}
for which a straightforward calculation gives:
\begin{eqnarray}
{\bar R} \!\!& = &\!\!
6 \, \frac{a''}{a^3}
\;\; , \label{barRconf1} \\
\Delta R \!\!& = &\!\!
6 \, \frac{a''}{a^3} \, h_{00} \, + \,
3 \, \frac{a'}{a^3} 
\Big[ \, h'_{00} - 2 h_{0i , i} + h'_{ii} \, \Big]
\nonumber \\
& \mbox{} &
+ \, \frac{1}{a^2} 
\Big[ \, \nabla^2 h_{00} - \nabla^2 h_{ii} - 
2 h'_{0i , i} + h''_{ii} + h_{ij , ij}
\, \Big]
\;\; , \label{deltaRconf1}
\end{eqnarray}
{\it (ii) the d'Alembertian perturbation} 
$\Delta \square \! :$
\begin{equation}
\square \; \equiv \; 
{\bar{\square}} \, + \, \Delta \square
\; \; , \label{deltabox}
\end{equation}
which directly follows from the definition 
(\ref{box}):
\begin{eqnarray}
{\bar{\square}} \!\!& = &\!\!
\frac{1}{\sqrt{-\bar{g}}} \;
\partial_{\mu} \Big( 
{\bar g}^{\mu\nu} \sqrt{-\bar{g}} \; 
\partial_{\nu} \Big)
\; = \;
\frac{1}{a^4} \, \partial_{\mu} \Big(
a^2 \, \eta^{\mu\nu} \, \partial_{\nu} \Big)
\nonumber \\
& = &\!\!
\frac{1}{a^2} \left[ \,
\nabla^2 - \partial^2_0 - 2 \frac{a'}{a} \, \partial_0
\, \right]
\; = \;
\frac{1}{a^2} \left[ \,
\nabla^2 - \frac{1}{a^2} \, \partial_0 \, 
a^2 \, \partial_0 \, \right]
\;\; , \label{barboxconf1} \\
\Delta \square \!\!& = &\!\!
\Delta \Big( \frac{1}{\sqrt{-g}} \Big) \,
\partial_{\mu} \Big[ 
{\bar g}^{\mu\nu} \sqrt{-\bar{g}} \; 
\partial_{\nu} \Big] \, + \,
\frac{1}{\sqrt{-\bar{g}}} \;
\partial_{\mu} \Big[ 
( \Delta g^{\mu\nu} ) \sqrt{-\bar{g}} \; 
\partial_{\nu} \Big]
\nonumber \\
& \mbox{} &
+ \, \frac{1}{\sqrt{-\bar{g}}} \;
\partial_{\mu} \Big[ 
{\bar g}^{\mu\nu} ( \Delta \sqrt{-\bar{g}} ) \; 
\partial_{\nu} \Big]
\label{deltaboxconf1} \\
\!\!& = &\!\!
\frac{1}{2a^2} \, \eta^{\mu\nu} \,
( \partial_{\mu} h ) \, \partial_{\nu} \, - \,
\frac{1}{a^4} \, \partial_{\mu} \Big[ 
a^2 \, h^{\mu\nu} \, \partial_{\nu} \Big]
\;\; . \label{deltaboxconf2}
\end{eqnarray}
As a result, the action of $\Delta \square$ on 
the background source $\bar{X}$ gives:
\begin{eqnarray}
(\Delta \square) \, {\bar X} \!\!& = &\!\!
\frac{1}{a^2} \, {\bar X}' \,
\Big[ - \frac12 h' - h'_{00 , 0} + h_{0i , i} \, \Big] 
\, - \,
h_{00} \, \frac{1}{a^4} \, \partial_0 \, 
a^2 \, \partial_0 \, {\bar X} 
\label{deltaboxXconf1} \\
& = &\!\!
\frac{1}{a^2} \, {\bar X}' \,
\Big[ - \frac12 h' - h'_{00 , 0} + h_{0i , i} \, \Big] 
\, + \,
{\bar R} \, h_{00} 
\;\; . \label{deltaboxXconf2}
\end{eqnarray}

Subtracting equation (\ref{deltaboxXconf2}) from
equation (\ref{deltaRconf1}) gives $\Delta X$ 
via (\ref{deltaX}) and, in turn,  $\Delta p$ via 
(\ref{deltapconf1}): 
\begin{eqnarray}
{\Delta p} \!\!& = &\!\! 
- \, \epsilon \Lambda^2 \, 
f'[- \epsilon \, {\bar X}] \times 
\frac{1}{\bar{\square}} \, \Bigg\{ \,
3 \, \frac{a'}{a^3} 
\Big[ \, h'_{00} - 2 h_{0i , i} + h'_{ii} \, \Big]
\nonumber \\
& \mbox{} &
+ \, \frac{1}{a^2} 
\Big[ \, \nabla^2 h_{00} - \nabla^2 h_{ii} - 
2 h'_{0i , i} + h''_{ii} + h_{ij , ij}
\, \Big]
\nonumber \\
& \mbox{} &
+ \, \frac{1}{a^2} \, {\bar X}' \,
\Big[ \; \frac12 h' + h'_{00 , 0} - h_{0i , i} 
\, \Big] \, \Bigg\} 
\;\; . \label{deltapconf2}
\end{eqnarray}

* {\it The deviation $\Delta u_0$} \\
It is important to note that $\Delta u_0$ can
be directly determined from the perturbation
of the 4-velocity timelike condition (\ref{u}):
\begin{eqnarray}
g^{\mu\nu} \, u_{\mu} u_{\nu} \; = \; - 1
& \Longrightarrow &\quad
(\Delta g^{\mu\nu}) \, {\bar u}_{\mu} {\bar u}_{\nu}
\, + \,
2 g^{\mu\nu} \, {\bar u}_{\mu} \, ({\Delta u}_{\nu})
\; = \; 0
\qquad \label{deltau0eqn1} \\
& \Longrightarrow &\quad
- h_{00} \, + \, \frac{2}{a} \, \Delta u_0
\; = \; 0
\;\; . \label{deltau0eqn2}
\end{eqnarray}
We trivially conclude that:
\begin{equation}
\Delta u_0 \; = \; \frac12 \, a \, h_{00}
\qquad , \qquad 
\Delta u^0 \; = \; \frac{1}{2a} \, h_{00}
\;\; . \label{deltau0}
\end{equation}
As a result, $\Delta T^{0}_{~0}$ -- given by 
equation (\ref{deltaT00conf1}) -- is simplified:
\begin{equation}
\Delta T^{0}_{~0} \; = \; 
- \Delta \rho
\;\; . \label{deltaT00conf2}
\end{equation}

\vspace{0.3cm}

{$\star \; $} {\it Identities} \\
In addition to the equations presented throughout the 
main text, various of the following expressions have 
been used to obtain the results of this subsection: 
\begin{eqnarray}
\sqrt{-g} \!\!& = &\!\!
\sqrt{-{\bar g}} \,
\left[ \, 1 + \frac12 \, h + \dots \, \right]
\quad , \quad 
h \; \equiv \; h^{\mu}_{~\mu} \; = \,
-h_{00} + h_{ii} 
\;\; , \qquad \label{IDsqrtg1} \\
{\bar g}_{\mu\nu} \!\!& = & \!\! 
a^2 \, \eta_{\mu\nu}
\quad , \quad
{\bar g}^{\mu\nu} \; = \; 
a^{-2} \, \eta^{\mu\nu}
\;\; , \label{IDbarg1} \\
\sqrt{-{\bar g}} \!\!& = &\!\! 
a^4
\quad , \quad
\frac{1}{\sqrt{-{\bar g}}} \; = \;
\frac{1}{a^4}
\;\; , \label{IDsqrtbarg1} \\
{\Delta g}_{\mu\nu} \!\!& = &\!\! 
a^2 \, h_{\mu\nu}
\quad , \quad
{\Delta g}^{\mu\nu} \; = \; 
-a^{-2} \, h^{\mu\nu}
\;\; , \label{IDdeltag1} \\
\Delta \sqrt{-g} \!\!& = &\!\! 
\frac12 \, a^4 \, h  
\quad , \quad
\Delta \Big( \frac{1}{\sqrt{-g}} \Big) \; = \; 
- \frac{1}{2 \, a^4} \, h
\;\; . \label{IDdeltasqrtg1} \\
\partial_{\mu} {\bar X} \!\!& = &\!\!
\delta^{0}_{~\mu} \, {\bar X}'
\;\; . \label{IDbarX1} 
\end{eqnarray}
When acting on functions that only depend on conformal
time:
\begin{eqnarray}
\bar{\square} \; = \;
- \frac{1}{a^4} \, \partial_0 \, a^2 \, \partial_0
\qquad , \qquad
\frac{1}{\bar{\square}} \; = \;
- \frac{1}{\partial_0} \, \frac{1}{a^2} \,
\frac{1}{\partial_0} \, a^4 
\;\; . \label{IDbarsquare1}
\end{eqnarray}

\vspace{0.3cm}

{$\bullet \;\; $} {\bf The Conservation Equations} \\
Of the conservation equations (\ref{cons2a}-\ref{cons2c}),
we shall use (\ref{cons2a}) and (\ref{cons2c}) to
express $\Delta \rho$ and $\Delta u_{\mu}$ in terms
of $\Delta p$. As a first step, we write them more
explicitly:
\begin{eqnarray}
\partial_{\mu} \Big[ \, 
\sqrt{-g} \; (\rho + p) \, u^{\mu} \, \Big]
\!\!& = &\!\!
\sqrt{-g} \; u^{\mu} \, \partial_{\mu} \, p
\;\; , \label{consA} \\
(\rho + p) \, u^{\mu}_{~ ; \nu} \, u^{\nu}
\!\!& = &\!\!
- \Big[ \, g^{\mu\nu} + u^{\mu} \, u^{\nu} 
\, \Big] \, \partial_{\nu} \, p 
\;\; , \label{consB}  
\end{eqnarray}
and study them up to first order. \\

{\it (i) Zeroth order} \\
Because the background 4-velocity is given by
(\ref{baruconf}), the background value of 
(\ref{consA}) equals:
\begin{equation}
\partial_0 \Big[ \, 
a^3 (\, {\bar{\rho}} + {\bar p} \,) \, \Big]
\; = \;
a^3 {\bar p}'
\;\; , \label{barconsAconf1}
\end{equation}
and integrates to:
\begin{equation}
{\bar{\rho}} + {\bar p} \; = \;
\frac{1}{a^3} \int_{\eta_0}^{\eta}
d \acute{\eta} \; a^3 \; {\bar p}'
\;\; , \label{barconsAconf2}
\end{equation}
which is the conformal analogue of equation
(\ref{rhofrw-pfrw}). The background value of 
(\ref{consB}) leads to a tautology of the
form $0=0$. \\

{\it (ii) First order} \\
When the perturbations defined in
(\ref{pertp2}-\ref{pertgmn}) are applied to the 
conservation equations (\ref{consA}-\ref{consB}) 
they respectively lead to:
\begin{eqnarray}
& \mbox{} &
\hspace{-1.7cm}
\partial_{\mu} \Big[ \, 
(\Delta \sqrt{-g}) \, (\bar{\rho} + {\bar p}) \, 
{\bar u}^{\mu} \, + \,
\sqrt{-{\bar g}} \, \Big( \Delta (\rho + p) \Big) \, 
{\bar u}^{\mu} \, + \,
\sqrt{-{\bar g}} \; (\bar{\rho} + {\bar p}) \, 
(\Delta u^{\mu}) \, \Big]
\nonumber \\
& \mbox{} &
\hspace{-0.7cm}
= \; (\Delta \sqrt{-g}) \, {\bar u}^{\mu} \, 
\partial_{\mu} \, {\bar p} \, + \,
\sqrt{-{\bar g}} \; (\Delta u^{\mu}) \, 
\partial_{\mu} \, {\bar p} \, + \,
\sqrt{-{\bar g}} \; {\bar u}^{\mu} \, 
\partial_{\mu} (\Delta p)
\;\; , \label{deltaconsA} 
\end{eqnarray}
\begin{eqnarray}
& \mbox{} &
\hspace{-0.9cm}
\Big( \Delta(\rho + p) \Big) \, 
{\bar u}^{\mu}_{~ ; \nu} \, {\bar u}^{\nu} \, + \,
({\bar{\rho} + {\bar p}) \, 
(\Delta  u}^{\mu}_{~ ; \nu}) \, {\bar u}^{\nu} \, + \,
(\bar{\rho} + {\bar p}) \, 
{\bar u}^{\mu}_{~ ; \nu} \, (\Delta u^{\nu})
\label{deltaconsB} \\
& \mbox{} &
= \;
- \Big[ \, (\Delta g^{\mu\nu}) + 
{\bar u}^{\mu} \, (\Delta u^{\nu}) +
(\Delta u^{\mu}) \, {\bar u}^{\nu} 
\, \Big] \, \partial_{\nu} \, {\bar p} \, - 
\Big[ \, {\bar g}^{\mu\nu} + 
{\bar u}^{\mu} \, {\bar u}^{\nu} 
\, \Big] \, \partial_{\nu} (\Delta p) 
\;\; . \nonumber   
\end{eqnarray}
Use of various background values and identities 
reduces equations (\ref{deltaconsA}-\ref{deltaconsB})
as follows:
\begin{equation}
\partial_0 \Big[ \, a^3 \, \Delta (\rho + p) \, \Big]
\; = \;
- \, a^4 (\bar{\rho} + {\bar p}) \Big[ \,
\partial_i \, \Delta u^i + \frac{1}{2a} \, h_{ii}'
\, \Big]
\, + \, a^3 \, \Delta p'
\;\; , \label{deltaconsA2}
\end{equation}

\vspace{-0.6cm}

\begin{eqnarray}
& \mbox{} &
\hspace{-1.2cm}
a^3 (\bar{\rho} + {\bar p}) \, \Big\{
(\Delta u^{\mu})' + 2 \, \frac{a'}{a} \, \Delta u^{\mu} - 
\frac{a'}{a} \, \delta^{\mu}_{~0} \, \Delta u^0 \, + \,
\frac{1}{a} \, \eta^{\mu\nu} \Big[ \,
h_{0\nu , 0} - \frac12 h_{00 , \nu} \, \Big] \Big\}
\label{deltaconsB2} \\
& \mbox{} &
= \; 
a^2 \, h^{0 \mu} \, {\bar p}' \, - 
\Big[ \, \Delta u^{\mu} + 
\delta^{\mu}_{~0} \, \Delta u^0 \, \Big]
a^3 \, {\bar p}' \, - \,
a^2 \, ( \eta^{\mu\nu} - 
\delta^{\mu}_{~0} \, \delta^{\nu}_{~0} ) \,
\partial_{\nu} (\Delta p)
\;\; . \nonumber
\end{eqnarray}
The last equation can be 3+1 decomposed and
further reduced:
\begin{eqnarray}
\left\{ \Big[ \, a^3 (\bar{\rho} + {\bar p}) 
\, \Big]^2 (\Delta u^0) \right\}'
\!\!& = &\!\!
\left\{ \Big[ \, a^3 (\bar{\rho} + {\bar p}) 
\, \Big]^2 \frac{1}{2a} \, h_{00} \right\}'
\;\; , \label{deltaconsB0} \\
\Big[ \, a^3 (\bar{\rho} + {\bar p}) \,
(a^2 \, \Delta u^i) \, \Big]'
\!\!& = &\!\!
- \Big[ \, a^4 (\bar{\rho} + {\bar p}) \, h_{0i} \, \Big]'
\, + \, 
\frac12 \, a^4 (\bar{\rho} + {\bar p}) \, \partial_i h_{00}
\nonumber \\
& \mbox{} &
- \, a^4 \, \partial_i (\Delta p)
\;\; , \label{deltaconsBi}
\end{eqnarray}
where we have extensively used (\ref{barconsAconf1}). 
Equation (\ref{deltaconsB0}) is trivially satisfied 
by the solution (\ref{deltau0}) for $\Delta u^0$. 
Given the perturbation $\Delta p$, equation 
(\ref{deltaconsBi}) determines $\Delta u^i$ up to its
arbitrary initial value. Then, equation (\ref{deltaconsA2}) 
has enough information at its disposal to determine 
$\Delta \rho$, again up to its arbitrary initial value.
Therefore, $\Delta T^{\mu}_{~\nu}$ has been completely 
specified to first order. \\ 

{$\star \; $} {\it Identities} \\
In addition to the equations presented throughout the 
main text, various of the following expressions have 
been used to obtain the results of this subsection: 
\begin{eqnarray}
{\bar u}^{\mu}_{~, \nu} \!\!& = &\!\!
- \frac{a'}{a^2} \, 
\delta^{\mu}_{~0} \, \delta^{0}_{~\nu}
\;\; , \label{IDbaru1} \\
{\bar u}^{\mu}_{~; \nu} \!\!& = &\!\!
{\bar u}^{\mu}_{~, \nu} \, + \,
{\bar{\Gamma}}^{\mu}_{~\nu\rho} \, {\bar u}^{\rho}
\; = \;
\frac{a'}{a^2} \, \Big[ \,
\delta^{\mu}_{~\nu} -
\delta^{\mu}_{~0} \, \delta^{0}_{~\nu} \, \Big]
\quad , \quad
{\bar u}^{\mu}_{~; \nu} \, {\bar u}^{\nu}
\; = \; 0
\;\; , \qquad \label{IDbaru2} \\
\Delta u_{\mu} \!\!& = &\!\!
(\Delta g_{\mu\nu}) \, {\bar u}^{\nu} +
{\bar g}_{\mu\nu} \, (\Delta u^{\nu})
\; = \;
a \, h_{\mu 0} +
a^2 \, \eta_{\mu\nu} \, (\Delta u^{\nu})
\;\; , \label{IDdeltau1} \\
\Delta u^{\mu} \!\!& = &\!\!
(\Delta g^{\mu\nu}) \, {\bar u}_{\nu} +
{\bar g}^{\mu\nu} \, (\Delta u_{\nu})
\; = \;
\frac{1}{a} \, h^{\mu 0} +
\frac{1}{a^2} \, \eta^{\mu\nu} \, (\Delta u_{\nu})
\;\; , \label{IDdeltau1b} \\
\Delta u^{\mu}_{~;\nu} \!\!& = &\!\!
\Delta u^{\mu}_{~,\nu} +
\bar{\Gamma}^{\mu}_{~\nu\rho} \, (\Delta u^{\rho}) +
(\Delta \Gamma^{\mu}_{~\nu\rho}) \, \bar{u}^{\rho} 
\;\; , \label{IDdeltau2} \\
(\Delta u^{\mu}_{~;\nu}) \, {\bar u}^{\nu} 
\!\!& = &\!\!
\frac{1}{a^2} \, \Big[ a \, (\Delta u^{\mu}) \Big]' \, + \,
\frac{1}{a^2} \, h^{\mu ~ '}_{~0} \, - \,
\frac{1}{2a^2} \, h_{00}^{~~ , \mu}
\;\; , \label{IDdeltau3} \\
\partial_{\mu} \, {\bar p} \!\!& = &\!\!
\delta^{0}_{~\mu} \; {\bar p}' \; = \;
\delta^{0}_{~\mu} \; 
\frac{1}{a^3} \Big[ \,
a^3 (\rho + p) \, \Big]'
\;\; . \label{IDbarp1}
\end{eqnarray}

\vspace{0.3cm}

{$\bullet \;\; $} {\bf The Full Equations} \\
At the cost of being redundant, we list the results 
derived in this section for the gravitational equations
of motion (\ref{eom3}) up to first order in the 
perturbations (\ref{pertp}-\ref{pertgmn}). \\
{\it (i) Background equations:}
We are given a background $FRW$ spacetime characterized 
by the scale factor $a$, and an induced background 
pressure ${\bar p} \,$; then: 
\begin{eqnarray}
(00) & \Longrightarrow & \quad
-3 \, \frac{{a'}^{\, 2}}{a^4} 
\; = \;
- \, 3H_0^2 \, + \, 
8 \pi G \Big[ \, {\bar p} - 
\frac{1}{a^3} \int_{\eta_0}^{\eta} 
d {\acute \eta} \; a^3 \, {\bar p}' \, \Big]
\;\; , \qquad \label{bar00eom1} \\
(0i) & \Longrightarrow & \quad
0 \; = \; 0
\;\; , \label{bar0ieom1} \\
(ij) & \Longrightarrow & \quad
\left[ -2 \, \frac{a''}{a^3} + \frac{{a'}^{\, 2}}{a^4} 
\, \right] \delta_{ij} 
\; = \;
- \, 3H_0^2 \, \delta_{ij} \, + \,
8 \pi G \, {\bar p} \, \delta_{ij}
\;\; . \label{barijeom1} 
\end{eqnarray}
{\it (ii) First order equations:}
We are given the spacetime perturbation $h_{\mu\nu}$
and the induced pressure perturbation $\Delta p \;$;
then:
\footnote{To avoid prohibitively long expressions 
we have not expressed all variables in terms of
$h_{\mu\nu}$ and $\Delta p$.}  
\begin{eqnarray}
(00) & \Longrightarrow & \quad
\Delta G^0_{~0} \, = \, 
8 \pi G \, ( - {\Delta \rho} ) 
\;\; , \label{delta00eom1} \\
(0i) & \Longrightarrow & \quad
\Delta G^0_{~i} \, = \, 
8 \pi G \; 
(\bar{\rho} + {\bar p}) \, 
\frac{1}{a} \, \Delta u_i
\;\; , \label{delta0ieom1} \\
(ij) & \Longrightarrow & \quad
\Delta G^{i}_{~j} \, = \, 
8 \pi G \; \Delta p \; \delta^i_{~j}
\;\; . \hspace{5.5cm} \label{deltaijeom1}
\end{eqnarray}

\vspace{0.3cm}

{$\bullet \;\; $} {\bf Scalar Perturbations} \\
The very nature of the gravitational equations 
of motion allows for scalar, vector and tensor 
perturbations. Of these, it is the scalar
perturbations that have, up to now, the biggest
phenomenological interest. A general scalar 
perturbation is characterized by four scalar 
functions $\phi, \; B, \; \psi, \; E$ which 
are defined as follows \cite{mukhanov1}:
\begin{eqnarray}
h_{00} \!\!& \equiv &\!\! 
-2 \phi
\;\; , \label{phi} \\ 
h_{0i} \!\!& \equiv &\!\! 
- B_{, i}
\;\; , \label{B} \\ 
h_{ij} \!\!& \equiv &\!\! 
-2 \psi \, \delta_{ij} \, - \,
2 E_{, ij}
\;\; , \label{psi,E} 
\end{eqnarray}
and which can be combined into two gauge invariant 
scalar field variables \cite{mukhanov1}: 
\begin{eqnarray}
\Phi \!\!& \equiv &\!\!
\phi \, - \, 
\frac{1}{a} \, \Big[ \, a \, (B - E') \, \Big]'
\; = \;
\phi \, - \, 
\frac{a'}{a} \, (B - E') \, - \,
(B' - E'')
\;\; , \qquad \label{Phi} \\
\Psi \!\!& \equiv &\!\!
\psi \, + \, \frac{a'}{a} \, (B - E')
\;\; . \label{Psi}
\end{eqnarray}

To deduce the equations that are obeyed by 
the scalar perturbations we must isolate their 
contribution to the full gravitational equations 
using (\ref{phi}-\ref{psi,E}) and then express 
it in terms of the invariant variables 
(\ref{Phi}-\ref{Psi}): \\

{\it (i) The left hand side} \\
The 3+1 decomposition of $(\Delta G^{\mu}_{~\nu})_{S}$
equals: 
\begin{eqnarray}
\Big( \Delta G^0_{~0} \Big)_{S} 
\!\!& = &\!\!
- \frac{2}{a^2} \, \nabla^2 \, \Big[ \,
\psi + \frac{a'}{a} (B - E') \, \Big] 
\, + \,
6 \, \frac{a'}{a^3} \, \Big[ \,
\psi' + \frac{a'}{a} \, \phi \, \Big]
\label{G00scalar1} \\
& = &\!\!
- \frac{2}{a^2} \, \nabla^2 \, \Psi
\, + \,
6 \, \frac{a'}{a^3} \, \Big[ \,
\Psi' + \frac{a'}{a} \, \Phi \, \Big]
\, + \,
{\bar G}^{0 ~ '}_{~0} (B - E')
\;\; , \label{G00scalar2} \\
\Big( \Delta G^0_{~i} \Big)_{S} 
\!\!& = &\!\!
- \, \frac{2}{a^2} \, \partial_i \Big[ \,
\psi' + \frac{a'}{a} \, \phi \, \Big]
\label{G0iscalar1} \\
& = &\!\!
- \, \frac{2}{a^2} \, \partial_i \Big[ \,
\Psi' + \frac{a'}{a} \, \Phi \, \Big]
\, + \,
\Big( {\bar G}^{0}_{~0} - 
\frac13 \, {\bar G}^j_{~j} \Big) \,
\partial_i (B - E')
\;\; , \label{G0iscalar2} \\
\Big( \Delta G^i_{~j} \Big)_{S} 
\!\!& = &\!\!
\delta_{ij} \times 
\frac{2}{a^2} \, \Bigg\{ \, \Big[ \,
2 \, \frac{a''}{a} - \frac{a'^{\, 2}}{a^2} \, \Big]
\, \phi \, + \,
\frac{a'}{a} \Big[ \, 2 \psi' + \phi' \, \Big]
\, + \, \psi''
\nonumber \\
& \mbox{} &
\hspace{1.3cm}
+ \, \frac12 \nabla^2 \Big[ \,
\phi - \psi - (B' - E'') - 2 \, \frac{a'}{a} (B - E')
\, \Big] \, \Bigg\}
\nonumber \\
& \mbox{} &
\hspace{-0.2cm}
+ \, \frac{1}{a^2} \, \partial_i \partial_j
\Bigg\{ \, \psi - \phi + B' - E'' + 
2 \, \frac{a'}{a} (B - E'') \, \Bigg\} 
\label{Gijscalar1} \\
& = &\!\!
\frac{2}{a^2} \, \Bigg\{ \,
\frac12 \nabla^2 (\Phi - \Psi) + \Psi'' + 
\frac{a'}{a} \, (2\Psi' + \Phi') + 
\Big[ \, 2 \, \frac{a''}{a} - 
\frac{a'^{\, 2}}{a^2} \, \Big] \Phi \, \Bigg\} 
\qquad \nonumber \\
& \mbox{} &
\times \, \delta_{ij} 
\, + \, \frac{1}{a^2} \, ( \Psi_{, ij} - \Phi_{, ij} )
\, + \, {\bar G}^{i ~ '}_{~j} (B - E')
\;\; . \label{Gijscalar2}
\end{eqnarray}

{\it (ii) The right hand side} \\
A similar analysis must be made to derive
$(\Delta T^{\mu}_{~\nu})_{S}$. We need only compute 
the scalar perturbations contribution to $\Delta p$ 
since the conservation equations will give the 
similar contribution to $\Delta \rho$ and $\Delta u_i$.
\footnote{The perturbation $\Delta u_0$ does not 
enter $(\Delta T^{\mu}_{\nu})_{S}$ and, at any rate, 
has been calculated: $\Delta u_0 = - a \phi$.}
Instead of calculating directly from (\ref{deltapconf2}), 
we shall perform the computation sequentially by first 
considering the scalar perturbations contribution to 
(\ref{deltaRconf1}):
\begin{eqnarray}
\Big( \Delta R \Big)_{S} \!\!& = &\!\!
- \, 12 \, \frac{a''}{a^3} \, \phi \, - \,
6 \, \frac{a'}{a^3} \, \Big[ \,
\phi' + 3 \psi' - \nabla^2 (B - E') \, \Big]
\nonumber \\
& \mbox{} &
+ \, \frac{2}{a^2} \, \Big[ \,
\nabla^2 (2 \psi - \phi) - 3 \psi'' +
\nabla^2 (B' - E'') \, \Big] 
\label{deltaRscalar1} \\
& = &\!\!
\frac{2}{a^2} \, \nabla^2 (2 \Psi - \Phi) \, - \,
\frac{6}{a^2} \, \Big[ \, \Psi'' +
\frac{a'}{a} \, (3 \Psi' + \Phi') \, \Big] 
\nonumber \\
& \mbox{} &
- \, 2 {\bar R} \, \phi \, + \,
\frac{1}{a^2} \, ( a^2 \, {\bar R} )' \,
(B - E') \, + \,
2 {\bar R} \, (B' - E'') 
\;\; , \qquad\qquad \label{deltaRscalar2}
\end{eqnarray}
and the similar part of (\ref{deltaboxXconf1}):
\begin{eqnarray}
\Big[ (\Delta \square) \, {\bar X} \Big]_{S}
\!\!& = &\!\!
- \, 2 {\bar R} \, \phi \, + \,
\frac{1}{a^2} \, {\bar X}' \, \Big[ \,
3 \psi' + \phi' - \nabla^2 (B - E') \, \Big]
\label{deltaboxXscalar1} \\ 
& = &\!\!
- \, 2 {\bar R} \, \phi \, + \,
\frac{1}{a^2} \, {\bar X}' \, \Bigg\{ \,
3 \Psi' + \Phi' - 
\Big[ \, 2 \, \frac{a''}{a} - 
2 \, \frac{a'^{\, 2}}{a^2} \, \Big] (B - E')
\qquad \nonumber \\
& \mbox{} & 
- \frac{2a'}{a} \, (B' - E'') + 
(B'' - E''') -
\nabla^2 (B - E') \, \Bigg\}
\;\; . \label{deltaboxXscalar2} 
\end{eqnarray}
By exploiting (\ref{barboxconf1}), the last term 
in (\ref{deltaboxXscalar2}) can be rewritten as 
follows:
\begin{eqnarray}
\frac{1}{a^2} \, {\bar X}' \, \nabla^2 
\Big[ \, (B - E') \, \Big] \!\!& = &\!\!
\frac{1}{a^2} \, \nabla^2 \; 
\Big[ \, {\bar X}' \, (B - E') \, \Big]
\nonumber \\
& \mbox{} &
\hspace{-3.5cm}
= \;
\bar{\square} \; 
\Big[ \, {\bar X}' \, (B - E') \, \Big]
\, + \,
\frac{1}{a^2} \Big( \,
\partial_0 + 2 \frac{a'}{a} \, \Big) \partial_0 \;
\Big[ \, {\bar X}' \, (B - E') \, \Big]
\qquad \label{id1} \\
& \mbox{} &
\hspace{-3.5cm}
= \;
\bar{\square} \; 
\Big[ \, {\bar X}' \, (B - E') \, \Big]
\, - \,
\frac{1}{a^2} \, ( a^2 \, {\bar R} )' \,
(B - E') \, - \,
2 {\bar R} \, (B' - E'') 
\nonumber \\
& \mbox{} &
\hspace{-2.5cm}
+ \, {\bar X}' \, \Bigg\{ \,
\frac{1}{a^2} \, (B'' - E''') \, - \,
2 \, \frac{a'}{a^3} \, (B' - E'') 
\nonumber \\
& \mbox{} &
- \, \Big[ \, 2 \, \frac{a''}{a^3} - 
2 \, \frac{a'^{\, 2}}{a^4} \, \Big]
(B - E') \, \Bigg\}
\;\; , \label{id2}
\end{eqnarray}
where in the second step we used another direct 
consequence of (\ref{barboxconf1}):
\begin{equation}
- \, \frac{1}{a^2} \Big( \,
\partial_0 + 2 \frac{a'}{a} \, \Big) \partial_0 \;
{\bar X} \; = \; {\bar R} 
\qquad  \Longrightarrow \qquad
{\bar X}'' \; = \,
- 2 \, \frac{a'}{a} \, {\bar X}' \, - \,
a^2 \, {\bar R}
\;\; , \label{id3} 
\end{equation}  
Appropriately combining equations (\ref{deltaRscalar2},
\ref{deltaboxXscalar2}, \ref{id2}), we get the scalar
perturbations part of the source deviation:
\begin{eqnarray}
\Big( \Delta X \Big)_{S} \!\!& = &\!\!
\frac{1}{\bar{\square}} \left[ \,
\Delta R -  (\Delta \square) \, {\bar X} 
\, \right]_{S}
\nonumber \\
& = &\!\!
{\bar X}' \, (B - E') \, + \,
\frac{1}{\bar{\square}} \, \Bigg\{ \,
\frac{2}{a^2} \, \nabla^2 (2 \Psi - \Phi) \, - \,
\frac{6}{a^2} \, \Big[ \, \Psi'' +
\frac{a'}{a} \, (3 \Psi' + \Phi') \, \Big] 
\nonumber \\
& \mbox{} &
\hspace{3.5cm}
- \, \frac{1}{a^2} \, {\bar X}' \, (3 \Psi' + \Phi')
\, \Bigg\} 
\;\; . \label{deltaXscalar1}
\end{eqnarray} 

The induced pressure first order scalar deviation 
follows from (\ref{deltapconf1}):
\begin{eqnarray}
{\Delta p} &\!\! = &\!\! 
{\bar p}' \, (B -E') \, - \, \epsilon \Lambda^2 \, 
f'[- \epsilon \, {\bar X}] \times 
\frac{1}{\bar{\square}} \, \Bigg\{ \,
\frac{2}{a^2} \, \nabla^2 (2 \Psi - \Phi) 
\nonumber \\
& \mbox{} &
\hspace{0.7cm}
- \, \frac{6}{a^2} \, \Big[ \, \Psi'' +
\frac{a'}{a} \, (3 \Psi' + \Phi') \, \Big] \, - \, 
\frac{1}{a^2} \, {\bar X}' \, (3 \Psi' + \Phi')
\, \Bigg\} 
\;\; . \label{deltapscalar1}
\end{eqnarray}

{\it (iii) The conservation equations} \\
The scalar perturbations part of the conservation
equations (\ref{deltaconsA2}, \ref{deltaconsB0},
\ref{deltaconsBi}) becomes respectively:
\footnote{We have decomposed the perturbation
$\Delta u_i$ into its transverse and longitudinal
parts: $\Delta u_i \equiv \Delta u_i^T +
\partial_i \, \Delta u$. The transverse perturbation
is of no concern to us.}
\begin{eqnarray}
& \mbox{} &
\hspace{-1cm}
\Big[ \, a^3 \, \Delta (\rho + p) \, \Big]'
\; = \; 
a^3 \, \Bigg\{ \Delta p' + 
(\bar{\rho} + \bar{p}) \Big[ \,
3 \psi' -  
\nabla^2 \Big( \frac{1}{a} \, \Delta u + B -E' \Big) 
\, \Big] \, \Bigg\}
\qquad \label{deltaconsA2scalar1} \\
& \mbox{} &
\left\{ \Big[ \, a^3 (\bar{\rho} + {\bar p}) 
\, \Big]^2 \, \Delta u^0 \right\}'
\; = \;
- \, \left\{ \Big[ \, a^3 (\bar{\rho} + {\bar p}) 
\, \Big]^2 \; \frac{1}{a} \, \phi \right\}'
\;\; , \label{deltaconsB0scalar1} \\
& \mbox{} &
\Big[ \, a^3 (\bar{\rho} + \bar{p}) \, 
\Delta u \, \Big]'
\; = \;
\Big[ \, a^4 (\bar{\rho} + \bar{p}) \, B \, \Big]'
\, - \, a^4 \Big[ \, \Delta p + 
(\bar{\rho} + \bar{p}) \, \phi \, \Big]
\;\; . \label{deltaconsBiscalar1} 
\end{eqnarray}
As before, equations (\ref{deltaconsA2scalar1},
\ref{deltaconsBiscalar1}) determine $\Delta u$ 
and $\Delta \rho$ while equation 
(\ref{deltaconsB0scalar1}) is automaticaly 
satisfied by $\Delta u_0 = - a \phi$. \\
 
{\it (iv) An observation} \\
Consider the scalar perturbations part of the 
$(ij)$ first order equation of motion:
\begin{equation}
\Big( \Delta G^i_{~j} \Big)_{S} 
\; = \;
8 \pi G \, \Big( \Delta T^{i}_{~j} \Big)_{S} \,
\; = \;
8 \pi G \, ( \Delta p ) \, \delta^i_{~j}
\;\; . \label{ijeomscalar1}
\end{equation}
When $i \neq j$, equation (\ref{ijeomscalar1})
simplifies considerably, since all terms 
proportional to $\delta_{ij}$ vanish, and 
allows us to conclude that $\Phi = \Psi$: 
\begin{equation}
(i \neq j)_{S} 
\quad \Longrightarrow \qquad
\frac{1}{a^2} \, ( \Psi_{, ij} - \Phi_{, ij} )
= 0
\qquad \Longrightarrow \qquad
\Phi = \Psi
\;\; , \label{Phi=Psi}
\end{equation}
where we have used equations (\ref{Gijscalar2},
\ref{deltapscalar1}) and the conformal time
derivative of the $(ij)$ background equation
of motion:
\begin{equation}
{\bar G}^{i ~ '}_{~j} 
\; = \;
8 \pi G \; {\bar T}^{i ~ '}_{~j}
\; = \;
8 \pi G \; {\bar p}' \, \delta^i_{~j} 
\; \; . \label{barijeomscalar1} 
\end{equation}

\vspace{0.3cm}

{$\star \; $} {\it Identities} \\
In addition to the equations presented throughout the 
main text, various of the following expressions have 
been used to obtain the results of this subsection: 
\begin{eqnarray}
h \!\!& = &\!\!
- h_{00} \, + \, h_{ii} \; = \;
2\, \phi \, - \, 6 \, \psi \, - \, 2 \, \nabla^2 E
\;\; , \label{IDh1} \\
h_{ii} \!\!& = &\!\! 
-6 \, \psi \, - \, 2 \, \nabla^2 E
\;\; , \label{IDh2} \\
h_{0i , i} \!\!& = &\!\! 
- \nabla^2 B
\;\; , \label{IDh3} \\ 
h_{ij , j} \!\!& = &\!\! 
-2 \, \psi_{, i} \, - \, 2 \, \nabla^2 E_{, i}
\;\; , \label{IDh4} \\
h_{ij , ij} \!\!& = &\!\!
-2 \, \nabla^2 \psi \, - \, 2 \, \nabla^4 E
\;\; , \hspace{8.2cm} \label{IDh5}
\end{eqnarray}

\vspace{-0.6cm}

\begin{eqnarray}
\Phi' \!\!& = &\!\!
\phi' \, - \, 
\left[ \, \frac{a''}{a} - \frac{a'^{\, 2}}{a^2}
\, \right] (B - E') \, - \,
\frac{a'}{a} \, (B' - E'') \, - \,
(B'' - E''')
\;\; , \qquad \label{IDPhi'} \\
\Psi' \!\!& = &\!\!
\psi' \, + \,
\left[ \, \frac{a''}{a} - \frac{a'^{\, 2}}{a^2}
\, \right] (B - E') \, + \,
\frac{a'}{a} \, (B' - E'') 
\;\; , \label{IDPsi'} \\
\Psi'' \!\!& = &\!\!
\psi'' \, + \,
\left[ \, \frac{a'''}{a} - 3 \, \frac{a' a''}{a^2} +
 2 \, \frac{a'^{\, 3}}{a^3}
\, \right] (B- E') 
\nonumber \\
& \mbox{} &
+ \,
\left[ \, 2 \, \frac{a''}{a} - 2 \, \frac{a'^{\, 2}}{a^2}
\, \right] (B'- E'') \, + \,
\frac{a'}{a} \, (B'' - E''') 
\;\; , \label{IDPsi''} \\
\Psi' \!\!\!& + &\!\!\! 
\frac{a'}{a} \, \Phi \; = \;
\psi' \, + \, \frac{a'}{a} \, \phi \, + \,
\left[ \, \frac{a''}{a} - \frac{a'^{\, 2}}{a^2}
\, \right] (B - E') 
\;\; , \label{IDPhiPsi1} \\ 
\Phi \!\!\!& - &\!\!\! 
\Psi \; = \;
\phi - \psi - (B' - E'') -
2 \, \frac{a'}{a} \, (B - E')
\;\; , \label{IDPhiPsi2} \\
3 \Psi' \!\!\!& + &\!\!\!
\Phi' \; = \; 
3 \psi' \, + \, \phi' \, + \, 
\left[ \, 2 \, \frac{a''}{a} - 
2 \, \frac{a'^{\, 2}}{a^2} \, \right] (B - E') 
+ \, \frac{2a'}{a} \, (B' - E'') 
\nonumber \\
& \mbox{} &
\hspace{1cm}
- \, (B'' - E''') 
\;\; , \label{IDPhiPsi3} \\
\Psi'' \!\!\!& + &\!\!\!
\frac{a'}{a} \, ( 3\Psi' + \Phi' )
\; = \; 
\psi'' \, + \, 
\frac{a'}{a} \, ( 3\psi' + \phi' ) \, + \,
\frac 16 \, (a^2 \, {\bar R})' \, (B - E') 
\nonumber \\
& \mbox{} &
\hspace{2.9cm}
+ \, \frac{a^2}{3} \, {\bar R} \, (B' - E'') 
\;\; , \label{IDPhiPsi4}
\end{eqnarray}

\vspace{-0.6cm}

\begin{eqnarray}
{\bar G}^{0 ~ '}_{~0} \!\!\!& = &\!\!\!
- \, 6 \, \frac{a'}{a^3} \left[ \,
\frac{a''}{a} - 2 \, \frac{a'^{\, 2}}{a^2} \, \right]
\;\; , \label{IDbarG1} \\
{\bar G}^{0}_{~0} \!\!\!& - &\!\!\! 
\frac13 \, {\bar G}^j_{~j} \; = \;
\frac{2}{a^2} \left[ \,
\frac{a''}{a} - 2 \, \frac{a'^{\, 2}}{a^2} \, \right]
\;\; , \hspace{6.5cm} \label{IDbarG2} \\
{\bar G}^{i ~ '}_{~j} \!\!& = &\!\!
- \, \frac{2}{a^2} \left[ \,
\frac{a'''}{a} - 4 \, \frac{a' \, a''}{a^2} + 
2 \, \frac{a'^{\, 3}}{a^3} \, \right]
\delta^{i}_{~j}
\;\; , \label{IDbarG3} \\
{\bar R}' \!\!& = &\!\!
6 \, \frac{a'''}{a^3} \, - \,
18 \, \frac{a' \, a''}{a^4}
\qquad , \qquad
(a^2 \, {\bar R})' \; = \;
6 \, \frac{a'''}{a} \, - \,
6 \, \frac{a' \, a''}{a^2}
\;\; , \label{IDbarR1} \\
{\bar p}' \!\!& = &\!\!
- \epsilon \Lambda^2 \, {\bar X}' 
f'[ - \epsilon \, {\bar X} ] 
\;\; , \hspace{7cm} \label{IDbarp2}
\end{eqnarray}

\vspace{-0.7cm}

\begin{eqnarray} 
\Delta u^i \!\!& = &\!\!
\frac{1}{a} \, B_{, i} \, + \,
\frac{1}{a^2} \, \Delta u_i
\;\; . \hspace{8.6cm} \label{IDdeltau4}
\end{eqnarray}

\vspace{0.3cm}

{$\bullet \;\; $} {\bf The Invariant Completions} \\
There is an elementary way to redefine the Einstein 
tensor perturbation $\Delta G^{\mu}_{~\nu}$ and the
stress-energy tensor perturbation 
$\Delta T^{\mu}_{~\nu}$ so that both sides of the
equations of motion are gauge invariant 
\cite{mukhanov1}:
\begin{equation}
\Delta G^{\mu}_{~\nu} \; = \;
8 \pi G \, \Delta T^{\mu}_{~\nu}
\qquad \longrightarrow \qquad
\Delta {\cal G}^{\mu}_{~\nu} \; = \;
8 \pi G \, \Delta {\cal T}^{\mu}_{~\nu}
\;\; . \label{inveom1}
\end{equation}
The idea is for the differences between 
$(\Delta G^{\mu}_{~\nu} \, , \,
\Delta T^{\mu}_{~\nu})$ and 
$(\Delta {\cal G}^{\mu}_{~\nu} \, , \, 
\Delta {\cal T}^{\mu}_{~\nu})$ to simultaneously
make the latter gauge invariant and obey the 
gravitational equations of motion. This can be 
accomplished in general with the redefinitions:
\begin{eqnarray}
\Delta {\cal G}^{0}_{~0} \!\!& = &\!\!
\Delta G^{0}_{~0} \, - \,
{\bar G}^{0 ~ '}_{~0} (B - E')
\;\; , \label{G00inv1} \\
\Delta {\cal G}^{0}_{~i} \!\!& = &\!\!
\Delta G^{0}_{~i} \, - \,
\Big( {\bar G}^{0}_{~0} - 
\frac13 \, {\bar G}^{j}_{~j} \Big)
(B - E')_{, i}
\;\; , \label{G0iinv1} \\
\Delta {\cal G}^{i}_{~j} \!\!& = &\!\!
\Delta G^{i}_{~j} \, - \,
{\bar G}^{i ~ '}_{~j} (B - E')
\;\; . \label{Gijinv1}
\end{eqnarray}   
For instance, in the case of scalar perturbations
these difference terms are precisely the only terms 
of $(\Delta G^{\mu}_{\nu})_{S}$ that are not gauge 
invariant as can be seen from (\ref{G00scalar2},
\ref{G0iscalar2}, \ref{Gijscalar2}). The corresponding
redefinitions of the stress-energy perturbation 
are obvious:
\begin{eqnarray}
\Delta {\cal T}^{0}_{~0} \!\!& = &\!\!
\Delta T^{0}_{~0} \, - \,
{\bar T}^{0 ~ '}_{~0} (B - E')
\;\; , \label{T00inv1} \\
\Delta {\cal T}^{0}_{~i} \!\!& = &\!\!
\Delta T^{0}_{~i} \, - \,
\Big( {\bar T}^{0}_{~0} - 
\frac13 \, {\bar T}^{j}_{~j} \Big)
(B - E')_{, i}
\;\; , \label{00eominv1} \\
\Delta {\cal T}^{i}_{~j} \!\!& = &\!\!
\Delta T^{i}_{~j} \, - \,
{\bar T}^{i ~ '}_{~j} (B - E')
\;\; . \label{Tijinv1}
\end{eqnarray}   

{\it (i) The scalar perturbations equations of 
motion} \\
Returning to the scalar perturbations that concern
us, we have:
\footnote{Since we are only concerned with scalar 
perturbations, from now on we shall not indicate 
this explicitly.}
\begin{eqnarray}
(00)_{S} & \Longrightarrow &
\Delta {\cal G}^{0}_{~0}
\; = \;
8 \pi G \, \Big\{ \,
\Delta T^{0}_{~0} \, - \, 
{\bar T}^{0 ~ '}_{~0} (B - E') \, \Big\}
\label{00inveom1} \\
& \Longrightarrow &
- \, \frac{2}{a^2} \, \nabla^2 \Phi 
\, + \, 6 \, \frac{a'}{a^3} \, \Phi' 
\, + \, 6 \, \frac{a'^{\, 2}}{a^4} \, \Phi
\; = \; 
8 \pi G \, ( - \Delta {\cal E} )
\;\; , \label{00inveom2} \\
(0i)_{S} & \Longrightarrow &
\Delta {\cal G}^{0}_{~i}
\; = \;
8 \pi G \, \Big\{ \,  
\Delta T^{0}_{~i} \, - \,
\Big( {\bar T}^{0}_{~0} - 
\frac13 \, {\bar T}^{j}_{~j} \Big)
(B - E')_{, i} \, \Big\}
\qquad \label{0iinveom1} \\
& \Longrightarrow &
- \, \frac{2}{a^2} \, \Phi'
\, - \, 2 \, \frac{a'}{a^3} \, \Phi
\; = \; 
8 \pi G \, a^{-1} ( \Delta {\cal U} )
\;\; ,\label{0iinveom2} \\
(ij)_{S} & \Longrightarrow &
\Delta {\cal G}^{i}_{~j}
\; = \;
8 \pi G \, \Big\{ \, 
\Delta T^{i}_{~j} \, - \,
{\bar T}^{i ~ '}_{~j} (B - E') \, \Big\}
\label{ijeominv1} \\
& \Longrightarrow &
\frac{2}{a^2} \, \Phi''
\, + \, 6 \, \frac{a'}{a^3} \, \Phi'
+ \Big[ \, 4 \, \frac{a''}{a^3} -
2 \, \frac{a'^{\, 2}}{a^4} \, \Big] \, \Phi
\; = \; 
8 \pi G \, ( \Delta {\cal P} )
\;\; . \label{ijinveom2}
\end{eqnarray}   
In arriving at equations (\ref{00inveom2},
\ref{0iinveom2}, \ref{ijinveom2}) -- besides the 
equality $\Phi = \Psi$ -- we used expressions
(\ref{G00scalar2}, \ref{G0iscalar2}, 
\ref{Gijscalar2}) respectively, as well as the 
definitions:
\footnote{We have again decomposed the perturbation
$\Delta {\cal U}_i$ into its transverse and longitudinal
parts: $\Delta {\cal U}_i \equiv \Delta {\cal U}_i^T +
\partial_i \, \Delta {\cal U}$.}
\begin{eqnarray}
\Delta {\cal E} \!\!& \equiv &\!\!
\Delta \rho \, - \, {\bar{\rho}}' (B - E')
\;\; , \label{calE} \\
\Delta {\cal U} \!\!& \equiv &\!\!
\Delta u \, + \, a \, (B - E')
\;\; , \label{calU} \\
\Delta {\cal P} \!\!& \equiv &\!\!
\Delta p \, - \, {\bar p}' \, (B - E')
\;\; . \label{calP1}
\end{eqnarray}
Our induced pressure {\it ansatz} provides 
$\Delta {\cal P}$:
\begin{equation}
\Delta {\cal P} \; = \;
- \, \epsilon \Lambda^2 \, 
f'[- \epsilon \, {\bar X}] \times 
\frac{1}{\bar{\square}} \, \Bigg\{ \,
\frac{2}{a^2} \, \nabla^2 \Phi \, - \, 
\frac{6}{a^2} \, \Big[ \, \Phi'' +
\frac{4a'}{a} \, \Phi' \, \Big] \, - \, 
\frac{4}{a^2} \, {\bar X}' \, \Phi'
\, \Bigg\} 
\label{calP2}
\end{equation}
The remaining two quantities $\Delta {\cal E}$,
$\Delta {\cal U}$ are obtained from the conservation 
equations (\ref{consAinv1}, \ref{consBinv1}). \\

{\it (ii) The initial value problem} \\
The appropriate set of initial value data for the 
system of equations of motion (\ref{00inveom2},
\ref{0iinveom2}, \ref{ijinveom2}) consists of the
following:
\begin{equation}
(\Delta {\cal E})_{\eta = \eta_0} \;\; \& \;\;
(\Delta {\cal U})_{\eta = \eta_0} : {\rm unrestricted}
\qquad , \qquad
(\Delta {\cal P})_{\eta = \eta_0} = 0
\;\; . \label{ivd} 
\end{equation}
The requirement $(\Delta {\cal P})_{\eta_0} = 0$
comes from the perfect fluid form (\ref{Tmn}) of 
the stress-energy tensor and the induced pressure
{\it ansatz} (\ref{p}) ; $\square^{-1}$ is the
retarded Green's function and vanishes on the
initial value surface. The initial value data
$(\Delta {\cal E})_{\eta_0}$,
$(\Delta {\cal U})_{\eta_0}$ are free and via
equations (\ref{00inveom2}, \ref{0iinveom2}) 
determine $\Phi_{\eta_0}$, $\Phi'_{\eta_0}$.
Then, equation (\ref{ijinveom2}) determines
$\Phi''_{\eta_0}$. \\

{\it (iii) The scalar perturbations conservation
equations} \\
The equations of motion are augmented by the
conservation equations (\ref{deltaconsA2scalar1},
\ref{deltaconsBiscalar1}) whose invariant
completions are: 
\begin{eqnarray}
& \Big[ &\!\!\!\!
a^3 \, \Delta {\cal E} \, \Big]'
\; = \;
- \, (a^3)' \, \Delta {\cal P} 
\, + \, 3 a^3 \, (\bar{\rho} + \bar{p}) \, \Phi'
\, - \, a^3 \, (\bar{\rho} + \bar{p}) \, 
\nabla^2 \Big( \frac{\Delta {\cal U}}{a} \Big)
\;\; , \qquad\quad \label{consAinv1} \\
& \Big[ &\!\!\!\!
a^3 \, (\bar{\rho} + \bar{p}) \, 
\Delta {\cal U} \, \Big]'
\; = \;
- \, a^4 \, \Delta {\cal P} 
\, - \, a^4 \, (\bar{\rho} + \bar{p}) \, \Phi
\;\; . \label{consBinv1}
\end{eqnarray}

\vspace{0.3cm}
 
{\it (iv) The dynamical content} \\
The fact that $(\Delta {\cal E})_{\eta_0} 
\, \& \, (\Delta {\cal U})_{\eta_0}$ are unrestricted 
and, therefore, $\Phi_{\eta_0} \, \& \, \Phi'_{\eta_0}$ 
are also unrestricted, implies that there is a scalar
degree of freedom which becomes dynamical due to the 
presence of our gravitationally induced stress-energy 
tensor $T^{\mu}_{~\nu}[g]$.

This is to be contrasted with the situation where the
origin of the stress-energy tensor is the matter sector 
of the theory; if we denote the relevant deviation of 
the latter tensor by $\Delta \Theta^{\mu}_{~\nu}$, 
we have:
\begin{eqnarray}
(00)_{S} & \Longrightarrow &\quad
- \, \frac{2}{a^2} \, \nabla^2 \Phi 
\, + \, 6 \, \frac{a'}{a^3} \, \Big[ \, 
\Phi' + \frac{a'}{a} \, \Phi \, \Big]
\; = \; 
8 \pi G \, \Delta \Theta^{0}_{~0}
\;\; , \label{00thetaeom1} \\
(0i)_{S} & \Longrightarrow &\quad
- \, \frac{2}{a^2} \, \partial_i \Big[ \,
\Phi' + \frac{a'}{a} \, \Phi \, \Big]
\; = \; 
8 \pi G \, \Delta \Theta^{0}_{~i}
\; = \; 
8 \pi G \, \partial_i (\Delta \Theta)
\;\; , \qquad \label{0ithetaeom1}
\end{eqnarray}
where the last equality in (\ref{0ithetaeom1})
is true because we consider scalar perturbations.
From (\ref{0ithetaeom1}) we immediately deduce:
\begin{equation}
\Phi' + \frac{a'}{a} \, \Phi
\; = \;
- \, \frac{a^2}{2} \, 8 \pi G \, 
\Delta \Theta
\;\; . \label{0ithetaeom2}
\end{equation}
Substituting (\ref{0ithetaeom2}) in the $(00)_{S}$
equation (\ref{00thetaeom1}):
\begin{equation}
- \, \frac{2}{a^2} \, \nabla^2 \Phi 
\, - \, 3 \, \frac{a'}{a} \,  
8 \pi G \, \Delta \Theta
\; = \; 
8 \pi G \, \Delta \Theta^{0}_{~0}
\;\; , \label{00thetaeom3}
\end{equation}
allows us to conclude that $\Phi$ has no dynamics
since we can solve for it from (\ref{00thetaeom3}). \\

{$\star \; $} {\it Identities} \\
In addition to the equations presented throughout the 
main text, various of the following expressions have 
been used to obtain the results of this subsection: 
\begin{eqnarray}
{\bar T}^{0 ~ '}_{~0} \!\!& = &\!\!
- \bar{\rho}' \; = \; 
- 3 \, \frac{a'}{a} \, (\bar{\rho} + \bar{p})
\;\; , \label{IDbarT1} \\
{\bar T}^{i ~ '}_{~j} \!\!& = &\!\!
{\bar p}' \, \delta^{i}_{~j}
\qquad , \qquad 
{\bar T}^{0}_{~0} \, - \, 
\frac13 \, {\bar T}^{j}_{~j} \; = \; 
(\bar{\rho} + \bar{p}) 
\;\; , \label{IDbarT2}
\end{eqnarray}

\section{Scalar Perturbations Equation Solutions}

In this section we will investigate solutions of 
the non-local evolution equation (\ref{ijinveom2}).
It is most convenient, for this purpose, to return
to co-moving coordinates and we shall do so:
\begin{eqnarray}
& \mbox{} &
\hspace{-1cm}
{\ddot \Phi} \, + \, 4H \, {\dot \Phi} \, + \,
(3H^2 + 2{\dot H}) \, \Phi \; =
\nonumber \\
& \mbox{} &
- \, \omega^2 \times
\frac{f'[- \epsilon \, {\bar X}]}{f'_{\rm cr}} \times 
\left( - \frac{1}{\bar{\square}} \right) \Bigg\{ \,
{\ddot \Phi} \, + \, 5H \, {\dot \Phi} \, + \,
\frac23 {\dot{\bar X}} \, {\dot \Phi} \, + \,
\frac{k^2}{3a^2} \, \Phi \, \Bigg\}
\;\; , \qquad \label{evo} 
\end{eqnarray}
where the critical point ${\bar X}_{cr}$ and oscillation
frequency $\omega$ are \cite{NctRpw2}:
\begin{equation}
1 \, - \, 8\pi \epsilon \, f[- \epsilon \, {\bar X}]
\; = \; 0 
\qquad , \qquad 
\omega^2 \, = \;
24\pi \, \epsilon^2 \Lambda \, f'_{cr}
\;\; . \label{crit,omega}
\end{equation}
The scalar d'Alembertian acting on a general function
equals:
\begin{equation}
- {\bar{\square}} \; = \;
\partial_t^2 \, + \, 3H \, \partial_t \, + \,
\frac{k^2}{a^2}
\;\; . \label{Box}
\end{equation}
It should be clear that equation (\ref{evo}) cannot
be solved exactly and we need to develop a methodology
that will allow us to extract the time evolution of the
scalar perturbations. \\

{$\bullet \;\; $} {\bf Strategy} \\
We can realize the inverse d'Alembertian using the 
mode functions $u(t,k)$ and $u^*(t,k)$ of the massless, 
minimally coupled scalar which obey the equations:
\begin{eqnarray}
\ddot{u}(t,k) \, + \, 3 H(t) \, \dot{u}(t,k) \, + \,
\frac{k^2}{a^2(t)} \, u(t,k) 
&\!\! = \!\!& 
0 
\;\; , \label{modeseqn} \\
u(t,k) \; \dot{u}^*(t,k) \, - \,
\dot{u}(t,k) \; u^*(t,k) 
&\!\! = \!\!& 
i \, a^{-3}(t) 
\;\; . \label{modeswronskian}
\end{eqnarray}
The solutions for general $a(t)$ are quite complicated 
\cite{NctRpw7} but we shall only require the asymptotic 
forms long before and long after first horizon crossing:
\begin{eqnarray}
k \gg H(t) a(t) & \Longrightarrow &\quad 
u(t,k) \; \approx \;
\frac1{\sqrt{2 k} \; a(t)} \, 
\exp \Biggl[ +ik \int_t^{+\infty} \! 
\frac{dt'}{a(t')} \Biggr] 
\;\; , \label{before1st} \\
k \ll H(t) a(t) & \Longrightarrow &\quad
u(t,k) \; \approx \;
\frac{H(t_k)}{\sqrt{2 k^3}} \,
\Biggl\{ \Bigl[ 1 + O(k^2) \Bigr] 
\nonumber \\
& \mbox{} & 
\hspace{2.3cm} 
+ \frac{i \, k^3}{H^2(t_k)} 
\Biggl[ \int_t^{+\infty} \! 
\frac{dt'}{a^3(t')} + O(k^2) \Biggr] \Biggr\} 
\;\; , \qquad \label{after1st}
\end{eqnarray}
where $t_k$ is the time of first horizon crossing:
\begin{equation}
k = H(t_k) \; a(t_k) 
\;\; . \label{1sthor}
\end{equation}

We can construct the retarded Green's function 
of $\bar{\square}$ using the mode functions 
$u(t,k)$ and $u^*(t,k)$. Hence the action of 
$\, -\square^{-1} \,$ on some function $f(t,k)$ 
gives:
\begin{eqnarray}
\Bigl( -\frac1{\bar{\square}} \Bigr) 
\Bigl\{ f \Bigr\}(t,k) 
&\!\! \equiv \!\!&
\nonumber \\
& \mbox{} &
\hspace{-2cm}
i \int_0^t \! dt' \; a^3(t') \Bigl[ \,
u(t,k) \; u^*(t',k) - u^*(t,k) \; u(t',k) \, \Bigr]
f(t',k)
\;\; , \label{invbox}
\end{eqnarray}
and further action on $\, -\bar{\square} f(t,k)$ only 
gives back the function $f(t,k)$ up to homogeneous 
solutions:
\begin{equation}
\Bigl( -\frac1{\bar{\square}} \Bigr) 
\Bigl\{ -\bar{\square} f \Bigr\}(t,k) 
\; = \;
f(t,k) \, + \, 
\alpha \; u(t,k) \, + \, \beta \; u^*(t,k) 
\;\; . \label{extract}
\end{equation}
The constants $\alpha$ and $\beta$ can be expressed 
in terms of the initial values of $f$ and its first 
derivatives:
\begin{eqnarray}
\alpha &\!\! = \!\!& 
i \Bigl[ \, \dot{u}^*(0,k) \; f(0,k) \, - \,
u^*(0,k) \; \dot{f}(0,k) \, \Bigr] 
\;\; , \label{alpha} \\
\beta &\!\! = \!\!& 
i \Bigl[ \, u(0,k) \; \dot{f}(0,k) \, - \,
\dot{u}(0,k) \; f(0,k) \, \Bigr] 
\;\; . \label{beta}
\end{eqnarray}

Our approach will be to divide time evolution into
successive regimes and obtain reliable approximate 
solutions to (\ref{evo}) within each regime. The choice
of these regimes is dictated by the actual physical
evolution of the system and by our desire to reliably
approximate (\ref{evo}). For instance, the last term 
can be irrelevant or important depending on whether 
the particular mode with wave number $k$ has or has 
not experienced first horizon crossing respectively. 
We shall assume that the wave number $k$ lies in the 
range for which the mode experiences first horizon 
crossing during inflation, but close enough to the 
end of inflation that the mode is at a cosmologically 
observable scale today. There are five epochs during 
which we seek approximate solutions to (\ref{evo}).

\begin{figure}
\centerline{\epsfig{file=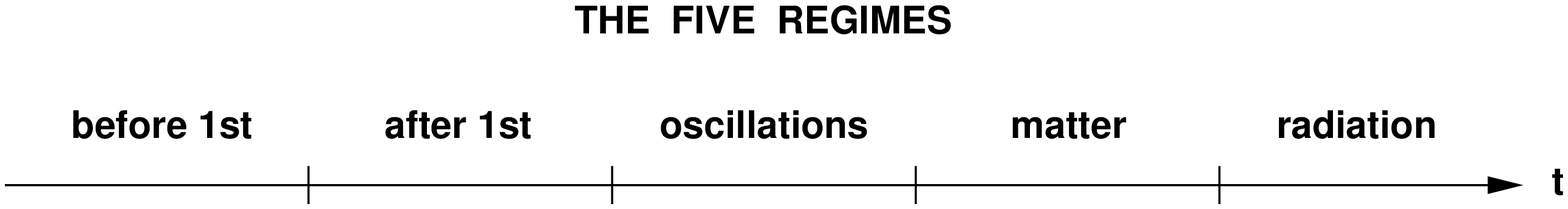,height=0.65in}}
\caption{\footnotesize The regimes during which the 
evolution of scalar perturbations was studied.}
\end{figure}

The strategy is always the same. We first determine 
the ``$f'$ ratio'' in the left hand side of (\ref{evo}),
then we approximate $\dot{\bar{X}}(t)$ in terms of 
$H(t)$. We next identify the ``large'' part of the 
curly bracketed term in the right hand side of (\ref{evo}) 
and extract it from the inverse d'Alembertian using 
(\ref{extract}), along with the appropriate homogeneous 
solution. \\

{$\bullet \;\; $} {\bf The Inflationary Regime Before 
First Horizon Crossing} \\
The simplest epoch is the first one during which:
\begin{equation}
H^2(t) \ll \frac{k^2}{a^2(t)} 
\;\; , \;\;
\omega^2 \ll H^2(t) 
\;\; , \;\;
\dot{\bar{X}}(t) \approx -4 H(t)
\;\; , \;\;
\frac{f'[-\epsilon \bar{X}(t)]}{f'[-\epsilon X_{cr}]} 
< 1
\label{regime1}
\end{equation}
and where the ``$f'$ ratio'' is very much less than one 
for most choices of the function $f(x)$. The ``large'' 
part of the curly bracketed term is $\frac{k^2}{3a^2}
\Phi$, so we extract it using (\ref{extract}, \ref{Box}):
\begin{eqnarray}
\left( - \frac{1}{\bar{\square}} \right) \Bigg\{ \,
{\ddot \Phi} \, + \, 5H \, {\dot \Phi} \, + \,
\frac23 {\dot{\bar X}} \, {\dot \Phi} \, + \,
\frac{k^2}{3a^2} \, \Phi \, \Bigg\}(t,k)
&\!\! \approx \!\!&
\nonumber \\
& \mbox{} & 
\hspace{-7.9cm} 
\frac13 \Phi(t,k) \, + \,
\Big( homogeneous \Big) \, + \,
\left( - \frac{1}{\bar{\square}} \right) \Bigg\{ \,
\frac23 \ddot{\Phi} \, + \, \frac43 H \, \dot{\Phi} 
\, \Bigg\}(t,k) 
\qquad\quad \label{evoregime1a}
\end{eqnarray}
where {\it (homogeneous)} stands for a linear 
combination of the homogeneous solutions $u(t,k)$ 
and $u^*(t,k)$ given by (\ref{before1st}). Because 
of conditions (\ref{regime1}) and the form of 
(\ref{before1st}), the entire right-hand side of 
(\ref{evo}) is negligible and the equation reduces 
to:
\begin{equation}
{\ddot \Phi} \, + \, 4H \, {\dot \Phi} \, + \,
(3H^2 + 2{\dot H}) \, \Phi 
\; \approx \; 0 
\;\; . \label{evoregime1b}
\end{equation}
Approximate solutions of (\ref{evoregime1b}) are:
\begin{equation}
\Phi_1 (t, k) \, \approx \, 
\frac{1}{a(t)}
\qquad , \qquad
\Phi_{1'} (t, k) \, \approx \, 
\frac{1}{a^3(t)}
\;\; , \label{regime1sols}
\end{equation}
and the recognition of this fact allows us also to 
estimate the first corrections from the non-local 
term. Suppose, for instance, we are correcting the 
$\Phi_1$ solution, in which case the non-local term 
is approximately given by the retarded Green's function 
(\ref{invbox}) -- constructed out of (\ref{before1st})
-- acting on the dominant term $\frac{k^2}{3a^2} \Phi_1$:
\begin{eqnarray}
\lefteqn{ 
\left( - \frac{1}{\bar{\square}} \right) \Bigg\{ \,
{\ddot \Phi} \, + \, 5H \, {\dot \Phi} \, + \,
\frac23 {\dot{\bar X}} \, {\dot \Phi} \, + \,
\frac{k^2}{3a^2} \, \Phi \, \Bigg\}(t,k) }
\nonumber \\
& \mbox{} & 
\hspace{1.1cm} 
\approx \;
\frac1{k \, a(t)} \int_0^t \! dt' \, a^2(t') \,
\sin \! \left[ k \int_{t'}^t \frac{dt''}{a(t'')} 
\right] \times 
\frac{k^2}{3 a^2(t')} \; \frac1{a(t')} 
\qquad \label{sol1corr1} \\
& \mbox{} & 
\hspace{1.1cm} 
= \; 
\frac1{3 a(t)} \int_0^t \! dt' \, \frac{k}{a(t')} \,
\sin \! \left[ k \int_{t'}^t \frac{dt''}{a(t'')} \right] 
\label{sol1corr2} \\
& \mbox{} & 
\hspace{1.1cm} 
= \;
\frac1{3 a(t)}
\left\{ 1 \, - \,
\cos \! \left[ k \int_{0}^t \frac{dt'}{a(t')} 
\right] \right\} 
\;\; . \label{sol1corr3}
\end{eqnarray}
The leading correction to $\Phi_1$ derives from the 
non-oscillating term $\frac1{3a}$ in (\ref{sol1corr3}).
However, this correction -- which grows like $\omega^2 
H^{-2} \ln a \times \Phi_1$ -- is of the same kind as 
the correction to $\Phi_1$ we would get if we take 
into account the $2 {\dot H} \Phi$ term in equation
(\ref{evoregime1b}).
\footnote{During inflation ${\dot H} \approx 
- \frac29 \omega^2$.}
Moreover, the dominant correction from the oscillating
term in (\ref{sol1corr3}) is, again, computed by inserting
it as the source in the evolution equation (\ref{evo}) 
and noting that the leading time dependence comes from 
the oscillating source term itself:
\begin{equation}
\Delta \Phi_1(t,k) \, = \,
\Phi_1(t,k) \, \times \, 
\frac{f'[-\epsilon \, \bar{X}(t)]}
{f'[-\epsilon \, X_{cr}]} \, \times 
\Big( -\frac13 \Big) \!
\left( \frac{\omega \, a(t)}{k} \right)^2
\cos \! \left[ k \int_{0}^t \frac{dt'}{a(t')} \right] 
\label{sol1corr4}
\end{equation}
The correction is very small long before first horizon 
crossing, owing to the factor of $\, k^{-2} \omega^2 
a^2(t)$. However, we see that the oscillatory term 
does grow. \\

{$\bullet \;\; $} {\bf The Inflationary Regime After 
First Horizon Crossing} \\
During this regime we are close enough to the end of 
inflation that, for instance, the ``$f'$ ratio'' is 
nearly unity: 
\begin{equation}
\frac{k^2}{a^2(t)} \ll H^2(t) 
\;\; , \;\;
\omega^2 \ll H^2(t) 
\;\; , \;\;
\dot{\bar{X}}(t) \approx -4 H(t)
\;\; , \;\;
\frac{f'[-\epsilon \bar{X}(t)]}{f'[-\epsilon X_{cr}]} 
\approx 1
\label{regime2}
\end{equation}
This is a difficult epoch to understand from first 
principles so we had recourse to explicit numerical 
studies. These revealed an end to the fall off in 
$\Phi(t,k)$ which characterizes the previous epoch.
In fact, the solution changes sign and its magnitude 
seems then to grow slowly. One can understand why 
this happens from the constant homogeneous solution 
which is built up by integrating the retarded Green's 
function over times $t'$ long before first horizon 
crossing. For this case we approximate $u(t,k)$ by 
expression (\ref{after1st}) and $u(t',k)$ by expression 
(\ref{before1st}). We further assume $\Phi(t',k) 
\approx a^{-1}(t')$, and that the $\frac{k^2}{3a^2} 
\Phi(t')$ term dominates:
\footnote{The contribution to the integration in
(\ref{sol2corr1}) from $t_k$ to $t$ is subleading
relative to that from $0$ to $t_k$.}
\begin{eqnarray}
\lefteqn{ 
\left( - \frac{1}{\bar{\square}} \right) \Bigg\{ \,
{\ddot \Phi} \, + \, 5H \, {\dot \Phi} \, + \,
\frac23 {\dot{\bar X}} \, {\dot \Phi} \, + \,
\frac{k^2}{3a^2} \, \Phi \, \Bigg\}(t,k) }
\nonumber \\
& \mbox{} & 
\approx \;
\frac{H(t_k)}{k^2} \int_0^{t_k} \! dt' \, a^2(t') \,
\sin \! \left[ k \int_{t'}^{+\infty} \frac{dt''}{a(t'')} 
\right] \times 
\frac{k^2}{3 a^2(t')} \; 
\frac1{a(t')} 
\qquad \label{sol2corr1} \\
& \mbox{} & 
= \; 
\frac1{3 a(t_k)} \, \left\{
\cos \! \left[ k \int_{t_k}^{+\infty} \frac{dt'}{a(t')} 
\right] \, - \,
\cos \! \left[ k \int_{0}^{+\infty} \frac{dt'}{a(t')} 
\right] \right\} 
\; \equiv \;
C_2(k)
\; \; , \qquad\quad \label{sol2corr2}
\end{eqnarray}
and we see that the constant is positive: 
$C_2(k) > 0$. Thus the approximate evolution 
equation during this epoch is:
\begin{equation}
{\ddot \Phi} \, + \, 4H \, {\dot \Phi} \, + \,
(3H^2 + 2{\dot H}) \, \Phi 
\; \approx \;
- \omega^2 \, C_2(k)
\;\; . \label{evoregime2}
\end{equation}
Even though $\, \omega^2 \ll H^2(t)$, the term on 
the right is not zero, so the $\, a^{-1}(t)$ fall 
off of $\Phi(t,k)$ cannot persist indefinitely. 
When this rapid time evolution of $\Phi(t,k)$ 
comes to an end the time derivative terms become 
insignificant and, because $\, -\dot{H}(t) \ll 
H^2(t)$, we have:
\begin{equation}
3 H^2(t) \, \Phi(t,k) \; \approx \; 
- \omega^2 \, C_2(k) 
\quad \Longrightarrow \qquad
\Phi(t,k) \, \approx \,
- \frac{\omega^2 \, C_2(k)}{3 H^2(t)} 
\;\; . \label{evosol2}
\end{equation}
Hence the solution changes sign and, because $H(t)$ 
decreases slowly during inflation, the magnitude of 
the solution increases slowly. That is exactly what 
the numerical simulations show. \\

{$\bullet \;\; $} {\bf The Oscillatory Regime } \\
During the epoch of oscillations the ``$f'$ ratio'' 
is still unity, and we also have:
\begin{equation}
\frac{k^2}{a^2(t)} \ll H^2(t) \ll 
\vert {\dot H} \vert \ll \omega^2
\;\; , \;\;
\dot{\bar{X}}(t) \approx -6 H(t)
\;\; , \;\;
\frac{f'[-\epsilon \bar{X}(t)]}{f'[-\epsilon X_{cr}]} 
\approx 1
\label{regime3}
\end{equation}
Expression (\ref{evosol2}) implies that $\Phi(t,k)$ 
must begin evolving again at the end of inflation,
so that its time derivatives are no long negligible. 
This has two consequences: first, the non-local term 
receives substantial contributions from times $t'$ 
after criticality; and second, the ``large'' term is
$\ddot{\Phi}$. Therefore we can write:
\begin{equation}
\left( - \frac{1}{\bar{\square}} \right) \Bigg\{ \,
{\ddot \Phi} \, + \, 5H \, {\dot \Phi} \, + \,
\frac23 {\dot{\bar X}} \, {\dot \Phi} \, + \,
\frac{k^2}{3a^2} \, \Phi \, \Bigg\}(t,k) 
\; \approx \;
\Phi(t,k) \, + \, C_3(k) 
\;\; , \label{sol3corr}
\end{equation}
where $C_3(k)$ consists of $C_2(k)$ plus a new, time 
independent contribution. Employing (\ref{regime3}),
the mode equation becomes effectively:
\begin{eqnarray}
\lefteqn{ \ddot{\Phi}(t,k) \, + \,
4 H(t) \, \dot{\Phi}(t,k) \, + \,
\omega^2 \, \Phi(t,k) 
\approx - \omega^2 \, C_3(k) } 
\label{evoregime3} \\
& \mbox{}& 
\hspace{1.5cm} 
\Longrightarrow \quad
\Phi(t,k) \; \approx \;
- C_3(k) \, + \, 
\frac{\Phi_3(k) \sin[\, \omega \Delta t + \phi_3(k) \,]}
{a^2(t)} 
\;\; , \qquad \label{evosol3}
\end{eqnarray}
where $\Delta t \equiv t - t_{\rm cr}$ and $\Phi_3(k)$,
$\phi_3(k)$ are constants. Had we included the first 
correction to the right hand side of (\ref{sol3corr}) 
the result would have been to slightly modify the rate 
of fall off in the oscillatory term, but it would not 
change the frequency. We emphasize that {\it all} the 
super-horizon modes oscillate at this same frequency. 
This is a profound distinction between the scalar 
perturbations of our model and those of scalar-driven 
inflation, and it has important consequences for 
reheating. \\

{$\bullet \;\; $} {\bf The Subsequent Matter and 
Radiation Domination Regimes} \\
Since the field $\Phi$ couples to matter universally
with gravitational strength, its oscillations with
frequency $\omega$ will most likely excite the particles
with masses $m \sim \omega$. These will be very heavy
particles which will behave -- even after their
excitation -- like non-relativistic matter. The very 
heavy particles that were the primary receptors of 
the energy from the oscillating field $\Phi$ will 
quickly decay into a hot radiation dominated universe. 

The analysis for these two epochs of matter is much the 
same. The only difference concerns the approximation 
we make for $\dot{\bar{X}}(t)$:
\begin{eqnarray}
matter & \Longrightarrow & \quad 
\dot{\bar{X}}(t) \; \approx \; -2 H(t) 
\;\; , \label{matterX} \\
radiation & \Longrightarrow & \quad
\dot{\overline{X}}(t) \; \approx \; 0 
\;\; . \label{radiationX}
\end{eqnarray}
These differences shows up only in the correction 
to (\ref{sol3corr}), which affects the rate of fall 
off but not the oscillatory frequency. Furthermore,
more evolution also affects the constant parts 
accumulated from the homogeneous solution:
\begin{eqnarray}
matter & \Longrightarrow & \quad
\Phi(t,k) \; \approx \;
- C_4(k) \, + \,
\frac{\Phi_4(k) \, 
\sin[\, \omega \Delta t + \phi_4(k) \,]}{a^2(t)} 
\;\; , \label{mattersol} \\ 
radiation & \Longrightarrow & \quad
\Phi(t,k) \; \approx \;
- C_5(k) \, + \,
\frac{\Phi_5(k) \, 
\sin[\, \omega \Delta t + \phi_5(k) \,]}{a^2(t)} 
\;\; . \qquad \label{radsol} 
\end{eqnarray}
We have {\it not} included the very significant 
decline in amplitude which must occur due to the 
flow of energy from the scalar modes into normal 
matter. It seems clear that this will continue 
until the amplitude of oscillation is driven to 
nearly zero. The final signal for the power spectrum 
resides in the constant $C_5(k)$ whose normalization
we cannot fix in the absence of canonical quantization.

\section{The Normalization of Perturbations}

Our effective field equations govern the time dependence 
of perturbations. We cannot actually solve these equations 
exactly but let us suppose, for the purposes of this 
discussion, that we could. For each wave vector $\bf{k}$ 
that would determine two linearly independent solutions,
$\Phi_1 (t, \bf{k})$ and $\Phi_2 (t, \bf{k})$. The full 
content of the effective field equations has been 
exhausted by expressing the perturbation operator 
$\widetilde{\Phi}(t, \bf{k})$ as a linear combination
of these two solutions:
\begin{equation}
\widetilde{\Phi}(t, \mathbf{k}) 
\; = \;
\alpha_1(\mathbf{k}) \times \Phi_1(t, \mathbf{k}) 
\; + \;
\alpha_2(\mathbf{k}) \times \Phi_2(t, \mathbf{k}) 
\;\; . \label{original}
\end{equation}
We can say what the operator coefficients, $\alpha_1(\bf{k})$ 
and $\alpha_2(\bf{k})$, are in terms of the initial values 
of $\widetilde{\Phi}(t, \bf{k})$ and its first derivative, 
but the field equations alone do not define how these 
operators commute. That information would ordinarily derive 
from applying canonical quantization to a Lagrangian, but 
in our non-local cosmological model we have no Lagrangian. 
We must instead regard the missing information as a separate 
assumption, which can be specified however we wish. It is 
a fundamental part of the definition of the model, every 
bit as much as the effective field equations were.

Before stating this assumption, let us clarify the 
issues in the very simple context of a 1-dimensional 
point particle whose position $q(t)$ obeys the simple
harmonic oscillator equation:
\begin{equation}
\ddot{q}(t) + \omega^2 q(t) \, = \, 0 
\;\; . \label{SHOeqn}
\end{equation}
This is a trivial equation to solve, and we can use it 
to express $q(t)$ in terms of its initial values $q_0$ 
and $\dot{q}_0$:
\begin{equation}
q(t) \; = \;
q_0 \, \cos(\omega t) \, + \,
\frac{\dot{q}_0}{\omega} \, \sin(\omega t) 
\;\; . \label{SHOsol}
\end{equation}
By decomposing the oscillatory functions into positive 
and negative frequencies we can identify linear combinations 
of the initial value operators which must lower and raise 
the energy:
\begin{eqnarray}
q(t) &\!\! = \!\!&
\frac12 \Big( q_0 + \frac{i\dot{q}_0}{\omega} \Big)
\, e^{-i \omega t} \, + \,
\frac12 \Big( q_0 - \frac{i\dot{q}_0}{\omega} \Big) 
\, e^{i \omega t} 
\nonumber \\
& \mbox{} &
\hspace{-2.4cm}
\Rightarrow \quad
\Big[ \, H \, , \, 
q_0 \pm \frac{i\dot{q}_0}{\omega} \, \Big] 
\; = \;
\mp \, \hbar \omega 
\Bigl( q_0 \pm \frac{i\dot{q}_0}{\omega} \Bigr) 
\; . \label{rlow}
\end{eqnarray}
Relation (\ref{rlow}) is as far as one can go using 
only the equation of motion (\ref{SHOeqn}). We do not 
know how $q_0$ and $\dot{q}_0$ commute, nor do we know 
how the Hamiltonian depends on them. Indeed, these two
issues are intimately related. If we ignore possible 
operator ordering ambiguities, the two Hamiltonian 
evolution equations:
\begin{eqnarray}
\dot{q}_0 &\!\! = \!\!& 
\frac{i}{\hbar} \, [H , q_0] 
\; = \;
-\frac{i}{\hbar} \,
\frac{\partial H}{\partial \dot{q}_0} \, 
[q_0 , \dot{q}_0] 
\;\; , \label{H1} \\
- \omega^2 q_0 &\!\! = \!\!& 
\frac{i}{\hbar} \, [H , \dot{q}_0] 
\; = \;
\frac{i}{\hbar} \,
\frac{\partial H}{\partial q_0} \, 
[q_0 , \dot{q}_0] 
\;\; , \label{H2}
\end{eqnarray}
are consistent with any solution of the form:
\begin{equation}
H = F(\mathcal{E}) 
\quad {\rm and} \quad 
[q_0 , \dot{q}_0] \, = \,
\frac{i\hbar}{F'(\mathcal{E})} 
\quad , \; {\rm where} \quad 
\mathcal{E} \, \equiv \,
\frac12 \dot{q}_0^2 \, + \,
\frac12 \omega^2 q_0^2 
\;\; . \label{F(E)}
\end{equation}
The equation of motion (\ref{SHOeqn}) cannot tell 
us what the function $F(\mathcal{E})$ is.

Note that there is still an ambiguity even if we 
assume $F(\mathcal{E})$ is linear -- which assumption 
might seem reasonable (athough not necessary) in view 
of the fact that the equation of motion is linear. 
The ambiguity rests with the proportionality constant: 
any function of the form $F(\mathcal{E}) = K \mathcal{E}$ 
would reproduce the canonical operator equations 
(\ref{H1}-\ref{H2}). 
\footnote{There is no operator ordering ambiguity for 
any linear ansatz.} 
Therefore, if we write $q(t)$ as a linear combination 
of canonically normalized creation and annihilation 
operators, the amplitude with which they appear contains
a factor of the arbitrary constant $K$:
\begin{equation}
q(t) \; = \;
\sqrt{\frac{\hbar}{2 K \omega}} \, \Bigl\{
a \, e^{-i\omega t} \, + \, 
a^{\dagger} \, e^{i\omega t} \Bigr\}
\quad , \quad 
[a , a^{\dagger}] \, = \, 1
\;\; . \label{K}
\end{equation}

It is canonical quantization of the simple harmonic 
oscillator Lagrangian which would ordinarily fix this 
constant:
\begin{equation}
L \; = \;
\frac12 m \, \dot{q}^2 \, - \,
\frac12 m \omega^2 q^2 
\qquad \Longrightarrow \qquad 
K = m 
\;\; . \label{K=m}
\end{equation}
In our case, the effective field equations are not associated 
with any Lagrangian. Indeed, the simultaneously causal and 
non-local nature of our equations precludes their derivation 
from any single-field Lagrangian. 
\footnote{Although a partial integration ``trick'' can
give causal non-local field equations, the resulting models
inevitably suffer an undesirable renormalization of the 
effective Newton's constant which the present class of
models avoids by construction \cite{ours}.}
A correct derivation from fundamental theory would involve 
an effetive action of the Schwinger-Keldysh type in which 
more than one quantity stands for what will eventually be 
the same dynamical variable, and one obtains the equation 
of motion by varying first and only then setting the 
different quantities equal \cite{SK}.

In the absence of an action principle we are forced to regard 
specification of the constant $K$ as an independent assumption, 
with the same status as the effective field equations themselves. 
This assumption can be made however we wish as part of how we 
define the model. In this context it should be noted that no 
principle seems to preclude the constant depending upon the 
magnitude of the wave vector, $k = \Vert \bf{k} \Vert$.
\footnote{Invariance under spatial rotations rules out more 
general dependence.} 
It is immediately obvious that we can enforce both the observed 
magnitude of scalar perturbations, and their approximate scale 
invariance, if only the super-horizon mode functions freeze in 
to constants in the late time regime. We saw in the previous 
section that the super-horizon mode functions do become constant 
during radiation domination, so that is the non-trivial check 
of our class of models, not either the magnitude nor the 
approximate scale invariance of scalar perturbations.

These comments need not apply to tensor perturbations. The 
equations which describe the latter are the same as those of 
general relativity, provided the expansion history $a(t)$ is 
fixed. We can therefore invoke the usual canonical normalization 
of the graviton creation and annihilation operators even though 
that is not, strictly speaking, required. Doing so makes the 
full tensor power spectrum a prediction of our model 
\cite{NctRpw5}.

\section{Reheating}

We have seen from Section 4 that the evolution of scalar 
perturbations reveals a profound difference between our 
model and any scalar-driven model of inflation:
\begin{itemize}
\item{In scalar-driven inflation the scalar mode functions 
oscillate and decay until horizon crossing, after which 
they approach constants; whereas}
\item{The scalar mode functions of our model decay until 
horizon crossing, then they become approximately constant 
until the end of inflation, after which they oscillate -- 
all with about the same frequency $\omega$ -- until enough 
energy has been dumped into the matter sector to support a
radiation dominated cosmology.}
\end{itemize}
Another key distinction is that, whereas no one knows or 
cares how long scalar-driven inflation persists beyond the 
50-60 e-foldings needed to solve the horizon and flatness 
problems, our model requires the vast number of 
$(G \Lambda)^{-1} \gtwid 10^6$ e-foldings of inflation. 
This implies that the number density $n$ of super-horizon 
modes at the end of inflation is staggering:
\begin{equation}
n \; \sim \;
\frac{H^3}{3 \pi^2} \, 
\exp \! \Big( \frac3{G \Lambda} \Big)
\; \gtwid \;
H^3 \times 10^{10^6} 
\;\; . \label{numden}
\end{equation}

Now consider what must happen when all of these modes start 
oscillating. There is not much energy in any one mode, and 
they couple only gravitationally to ordinary matter, so energy 
flow from them is very weak. However, the number density 
(\ref{numden}) of modes which participate is so enormous that 
reheating must be practically instantaneous.

It is interesting to contrast reheating in our model with 
the way it works in traditional single-scalar inflation. 
In those models the single inflaton zero mode undergoes 
oscillations at the end of inflation. At this stage it
becomes necessary to assume that the inflaton couples to 
ordinary matter because having the its kinetic energy 
transferred to gravity and, then, from gravity to ordinary 
matter, would be far too slow to reheat the universe. What 
would happen instead is that the energy would be reshifted 
away before it had a chance to accumulate and thermalize.

Direct couplings between the inflaton and ordinary matter 
can result in significant reheating. However, it comes at 
the price of more fine tuning because such couplings cause 
loops of matter quanta to induce Coleman-Weinberg terms in 
the inflaton effective potential $V_{\rm eff}$. These 
contributions would not be in terms of naturally small 
parameters, they would necessarily involve the relatively 
large coupling constants of ordinary matter. For example, 
a cubic coupling between the inflaton $\varphi$ and either 
a spectator scalar $\chi$ or a fermion field $\psi$ would 
induce \cite{CW}:
\begin{eqnarray}
- \, g_{\chi} \, \varphi \chi^2 
& \Longrightarrow & \quad
\Delta V_{\rm eff} \; = \;
+ \, \frac{g_{\chi}^2 \, \varphi^4}{16 \pi^2} \, 
\ln \Bigl( \frac{\varphi^2}{\mu^2} \Bigr) 
\;\; , \label{scalars} \\
- \, g_{\psi} \, \varphi \, \overline{\psi} \psi 
& \Longrightarrow & \quad
\Delta V_{\rm eff} \; = \;
- \, \frac{g_{\psi}^2 \, \varphi^4}{16 \pi^2} \, 
\ln \Bigl( \frac{\varphi^2}{\mu^2} \Bigr) 
\;\; , \label{fermions}
\end{eqnarray}
for some renormalization scale $\mu$. Either of these 
contributions would render the inflaton potential far 
too steep for successful inflation; the fermionic 
contribution (\ref{fermions}) would actually make the 
inflaton unstable. Hence these effects must be canceled 
by adding carefully chosen terms to the classical action. 
As long as one is not restricted to renormalizable 
inflaton potentials this can be done, but it represents 
a new level of fine tuning.

This additional challenge for scalar-driven inflation 
derives from the fact that reheating draws its energy 
from the oscillations of just a single scalar zero mode. 
Thus, the inflaton must be directly coupled to ordinary 
matter to give efficient reheating. By contrast, the 
reheating in our model draws its energy from the vast 
reservoir of super-horizon modes which are naturally 
accumulated during the long epoch of inflation. Because 
so many modes participate, it is not necessary (or even 
possible) to introduce a new, direct coupling to ordinary 
matter; gravitational couplings will suffice.

\section{Epilogue}

The phenomenological model considered in this study
is solely based on the graviton and therefore, one
would argue, it should have major problems in 
reproducing {\it any} realistic scalar density 
perturbations; after all, the graviton is a tensor
field possessing four unconstrained initial value
data which result in its two physical polarizations.
Unless we are willing to invoke graviton bound states,
there is simply no physical scalar degree of freedom 
present. However, this argument ignores the presence
of the gravitationally induced non-local source term
in the field equations. As we showed in Section 3,
its presence changes the dynamical content of the 
theory and, besides the two graviton polarizations,
the scalar $\Phi$ emerges as a physical degree of 
freedom possessing two unconstrained initial value 
data: $\Delta {\cal E}_0$ and $\Delta {\cal U}_0$ .

\begin{figure}
\centerline{\epsfig{file=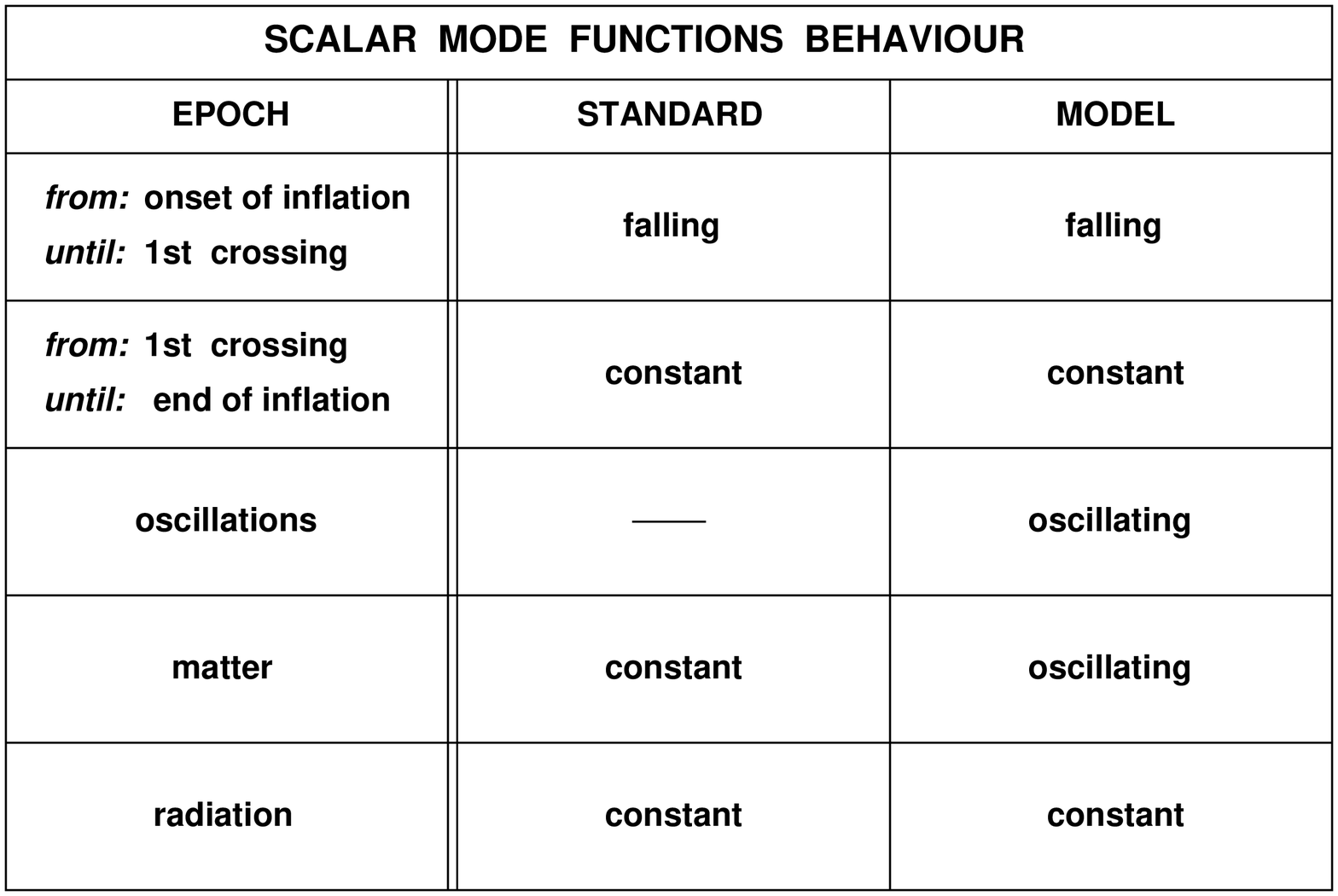,height=3.4in}}
\caption{\footnotesize The mode functions evolution 
in scalar-driven {\it vs} gravity-driven cosmology.}
\end{figure}

Deriving the evolution equation for the scalar $\Phi$ 
is a non-trivial exercise described in detail in 
Section 3. The (approximate) solutions to the equation 
in five successive regimes of evolution from the onset 
of inflaton until late times, showed distinctive 
differences with the standard inflationary picture; 
they are qualitatively recorded in Figure 2. What
counts is agreement with measurements and what is
measured is correlations between different portions
of the sky. These correlations are non-zero for the
scalar part of the generic perturbations in the
gravitational theory. Scalar-driven inflation makes
a non-zero contribution dictated by the form of the
inflaton Lagrangian. Our gravity-driven model also
makes a non-zero contribution which -- with the proper
normalization choice -- is consistent with the
observed magnitude and approximate scale invariance
of the scalar spectrum.

The novel feature of the gravity-driven model is 
the presence of an era subsequent to inflation 
during which {\it all} modes of $\Phi$ oscillate. 
By all modes we mean all infrared modes since
$\Phi$ is a dynamical degree of freedom emerging
in the infrared sector of the theory. Now the 
``receiver'' of the energy generated by the 
oscillations will be the matter sector of the 
theory. If radiation domination is reached -- and 
our model is predisposed to do so \cite{NctRpw6}
because radiation is the unique power law solution 
for which our simple source vanishes ($R = 0$) --
the energy deposited into matter will sustain
the radiation domination.

Each mode will contribute a very small amount of
energy to the process but since there exist a huge
amount of modes the whole process can be very 
efficient. Moreover, this reheating mechanism is 
very natural because the interaction of the modes
with any matter field is of the {\it universal}
gravitational strength. In contradistinction,
scalar-driven theories involve different couplings 
of the inflaton to different matter fields. In
one sentence, the scalar $\Phi$ via its coherent
oscillations can naturally reheat and lead to a 
hot thermal universe.

The simple phenomenological model used in this 
paper has late time evolution problems which can 
be addressed by modifying the {\it ansatz} for
the gravitationally induced source \cite{NctRpw6}.
This change, however, only affects the late time 
evolution and does not disturb the results of this 
paper concerning primordial scalar perturbations.

\vspace{1cm}

\centerline{\bf Acknowledgements}
This work was partially supported by the European 
Union grant FP-7-REGPOT-2008-1-CreteHEPCosmo-228644, 
by the NSF grants PHY-0653085 and PHY-0855021, and 
by the Institute for Fundamental Theory at the 
University of Florida.

\end{document}